\documentstyle[prl,aps,preprint,tighten,floats,aps,amssymb]{revtex}
\input epsf.tex

\newcommand{\bmat}{\left(\begin{array}}
\newcommand{\emat}{\end{array}\right)}
\def\NPB#1#2#3{Nucl. Phys. B{#1} (19#2) #3}
\def\PLB#1#2#3{Phys. Lett. B{#1} (19#2) #3}

\def\PRD#1#2#3{Phys. Rev. D{#1} (19#2) #3}

\def\yzero{\smash{\hbox{$y\kern-4pt\raise1pt\hbox{${}^\circ$}$}}}
\def\p{\partial}

\def\g{\gamma}

\def\-{\hphantom{-}}
\def\ov{\overline}
\def\s2{\frac{1}{\sqrt2}}

\def\beq{\begin{equation}}
\def\eeq{\end{equation}}
\def\beqa{\begin{eqnarray}}
\def\eeqa{\end{eqnarray}}

\def\Tr{{\rm Tr \,}}
\def\diag{{\rm diag \,}}

\def\IF{\relax{\rm I\kern-.18em F}}
\def\II{\relax{\rm I\kern-.18em I}}
\def\IP{\relax{\rm I\kern-.18em P}}

\def\Dsl{\,\raise.15ex\hbox{/}\mkern-13.5mu D} %can be subscripted

\def\IC{\bf C}
\def\IZ{\bf Z}

\def\z2z2{$\IC^3/(\IZ_2\times\IZ_2)$}

\def\id{{\bf 1}}
\def\Dthree{${\ov {\rm D3}}$}

\def\g{\gamma}

\def\p{\pi}
\def\q{\theta}

\def\s{\sigma}

\def\z{\zeta}

\def\O{\Omega}

\def\bo{{\raise-.3ex\hbox{\large$\Box$}}}               % D'Alembertian
\def\face{{\raise.2ex\hbox{$\displaystyle \bigodot$}\mskip-2.2mu \llap {$\ddot
        \smile$}}}                                      % happy face
\def\leftrightarrowfill{$\mathsurround=0pt \mathord\leftarrow \mkern-6mu
        \cleaders\hbox{$\mkern-2mu \mathord- \mkern-2mu$}\hfill
        \mkern-6mu \mathord\rightarrow$}       % <--> double differential
\def\dvec#1{\vbox{\ialign{##\crcr
        \leftrightarrowfill\crcr\noalign{\kern-1pt\nointerlineskip}
        $\hfil\displaystyle{#1}\hfil$\crcr}}}           % <--> accent
\def\beq{\begin{equation}}
\def\eeq{\end{equation}}

\def\beqx{\begin{displaymath}}
\def\eeqx{\end{displaymath}}

\def\beqa{\begin{eqnarray}}
\def\eeqa{\end{eqnarray}}

\begin{document}
\draft
\date{\today}
\title{
\normalsize
\mbox{ }\hspace{\fill}
\begin{minipage}{7cm}
UPR-904-T,\\
 CERN-TH/2000-270, \\
FERMILAB-Pub-00/257-T,\\
RUNHETC-2000-36,\\
SISSA-Ref.89/2000/EP \\
{\tt hep-th/yymmxxx}{\hfill}
\end{minipage}\\[5ex]
{\large\bf Discrete Wilson Lines in N=1 D=4  Type IIB Orientifolds: 
\\
 A Systematic Exploration for $\IZ_6$  Orientifold\\[1ex]}}

\author{ Mirjam Cveti\v c$^1$,  Angel M. Uranga$^2$ and Jing Wang$^3$} 
\address{$^1$Dept. of Physics and Astronomy, 
Univ. of Pennsylvania, Philadelphia PA 19104-6396, USA;\\
Dept. of Physics and Astronomy, Rutgers Univ., Piscataway,  NJ 08855-0849,
USA;\\
SISSA-ISAS and INFN, Sezione di Trieste, Via Beirut 2-4, I-34013, Trieste, Italy\\
$^2$ Fermilab, Theory Division, Batavia, IL 60510, USA\\
 $^3$  Theory Division, CERN, CH-1211 Geneva 23, Switzerland
}

\maketitle

\thispagestyle{empty}

\begin{abstract}
We develop techniques to construct general discrete Wilson lines in 
four-dimensional N=1 Type IIB  orientifolds, their T-dual realization 
corresponds to branes positioned at the orbifold fixed points. The explicit   
order two and three Wilson lines along with their tadpole consistency 
conditions are given for D=4 N=1 $\IZ_6$ Type IIB orientifold. The 
systematic search for all models with general order three Wilson
lines leads to a small class of inequivalent models. There are only two
inequivalent classes of a potentially phenomenologically interesting
model that has
a possible $SU(3)_{color}\times SU(2)_L\times SU(2)_R\times U(1)_{B-L}$
gauge structure, arising from a set of branes located at the
$\IZ_6$ orbifold fixed point. We calculate the spectrum and Yukawa couplings for this model. On the other hand, introduction of
anti-branes allows for models with three families and realistic gauge
group  assignment, arising from branes located at the $\IZ_3$ orbifold
fixed points.
 
\end{abstract}
%\pacs{\tt PACS number(s): }
%\vskip2cm
\newpage

\section{Introduction}
Four-dimensional N=1 supersymmetric Type IIB orientifolds (\cite{ABPSS,berkooz,N1orientifolds,kakuz6,ibanez,afiv,TyeKakush,wilsonlinemodel,CPW} and references therein) provide a fertile ground to study
new physics implications of perturbative open-string theories which in
turn could shed light on the physics of strongly coupled heterotic string
models. In this effort one is at fairly early stages. One goal, that is far 
from being achieved, is the development of techniques that would yield a
larger class of solutions (see e.g., \cite{asym} and references therein)
than those based on symmetric orientifold constructions. Another
direction is the exploration of possible deformations of symmetric
orientifold models, which in turn may provide large new classes of models
which may eventually lead to models with quasi-realistic features.  
 
Such deformations fall into two classes: (i) Wilson lines, which in the
T-dual picture correspond to branes moved to different positions on the  
orientifold, (ii) blow-up of orientifold singularities.  
  
The deformations in the space of  supersymmetric four-dimensional solutions
have a  field-theoretic realization; one identifies specific D- and F-flat 
directions within the effective theory of the original model \footnote{Such 
techniques have been extensively applied to the studies of heterotic string 
models, see. e.g., \cite{CCEELW} and references therein.}. The effective
field theory at the new (deformed) supersymmetric ground states is a 
power-series in the magnitude of the vacuum expectation values of the
fields responsible for the deformation. In particular, this technique was
used to study the blowing-up of the orientifold singularities
\footnote{Previous work on blowing up orbifold singularities has appeared 
in \cite{dgm}.} \cite{CELW,pru}, where the $\IZ_3$ \cite{ABPSS} and
$\IZ_2\times \IZ_2$ N=1 D=4 orientifolds are used as prototypes. Results are
different in nature from those of perturbative heterotic orbifolds \cite{C}.
The blowing-up deformation is notoriously difficult to describe within the
full string theory context, since the metric of the blown-up space is not
explicitly known. On the other hand the explicit field-theoretic realization 
in terms of (non-Abelian) flat directions allows for  the determination
of
the surviving gauge groups and massless spectrum.

On the other hand Wilson lines, both continuous and discrete, can in 
principle be constructed within full string theory. The T-dual interpretation
of continuous Wilson lines corresponds to a set of branes located at
generic points on the orientifold, while discrete Wilson lines correspond to 
special cases of sets of branes located at the orbifold fixed points.  
Deformation by continuous Wilson lines admits an interpretation in the
effective field theory as specific D- and F-flat directions, which
describe (in the T-dual picture) infinitesimal motion of sets of branes away 
from the fixed points.
  
Explicit examples of discrete Wilson lines have been constructed for a
number of different orientifold models (see \cite{afiv,wilsonlinemodel,CPW,lykken} and references therein). Continuous Wilson lines were first
addressed in \cite{ibanez,afiv}, but the explicit unitary representation
was not given. The explicit unitary representation of the continuous
Wilson line, specifically constructed for the $\IZ_3$ orientifold, along
with the most general set of continuous Wilson lines and their field
theory realization for $\IZ_3$ orientifold was recently given in
\cite{cl}.
 
The purpose of this paper is to advance the techniques to construct models 
with general discrete Wilson lines for general symmetric Type IIB  
orientifolds. In particular we construct the explicit form of such Wilson
lines which satisfy both the algebraic consistency conditions and 
the tadpole consistency conditions. We carry out our discussion in
 the T-dual picture, where
Wilson lines have a geometric interpretation as sets of branes located at
different fixed points. Algebraic consistency conditions amount to the
consistency of the location of these branes with the orbifold/orientifold
symmetries. We also show that the T-dual tadpole consistency conditions are
related to the original ones by a discrete Fourier transform. We apply
these techniques to fully explore the $\IZ_6$ orientifold model with a
very general choice of discrete Wilson lines. However, we would like to
emphasize the techniques are general, and apply to other orientifolds as
well.
 
One of the motivations to study this particular model is that it could be
very interesting from the phenomenological point of view, since it may
allow for putting branes at $\IZ_3$ {\em orbifold} points (that are not
fixed under orientifold projection). The relevance of configurations of
branes at $\IZ_3$ orbifold points to obtain potentially realistic models
has been discussed in \cite{Aldazabal}. If six branes are located at such
points such sectors may provide standard model gauge groups with three
families. However, we find the consistency constraints do not allow
for the gauge structure $SU(3)_{color}\times
SU(2)_L\times U(1)$ in such sectors. This is only possible if one includes 
anti-branes in the model and breaks supersymmetry (in a hidden sector).
 
Nevertheless we carry out a systematic exploration of all the possible
models with N=1 supersymmetry and find a small class of inequivalent models.
When restricting to models containing sectors with $SU(3)$ gauge group
factors, as candidates for $SU(3)_{color}$, we find only two inequivalent
classes, with a possible gauge group assignment $SU(3)_{color}\times
SU(2)_L\times SU(2)_R\times U(1)_{B-L}$ that arises from branes located at 
the $\IZ_6$ orbifold fixed point.
We provide the full massless spectrum and the Yukawa couplings, thus
setting the stage for further phenomenological explorations of this model.

The paper is organized in the following way. In Section \ref{wl} we construct 
both order three (Subsection \ref{wlo3}) and order two
(Subsection \ref{wlo2}) 
discrete Wilson lines  for $\IZ_6$ orientifold, as well as continuous Wilson 
lines (Subsection \ref{cwl}), and provide their T-dual interpretation. 
In Section \ref{tadpole} we provide the tadpole cancellation conditions in
a class of models with arbitrary number of order three Wilson lines.
For completeness we also describe the tadpole constraints in the T-dual
version, and show (in the Appendix) they are related to the tadpole
conditions 
in the original picture by a discrete Fourier transform. In Section
\ref{spectrum} we describe in detail the orbifold and orientifold projections, 
both in terms of D9- and D5$_3$-branes \footnote{Throughout this paper, 
we denote by D5$_i$-branes (D7$_i$-branes) the D5-branes (D7-branes) wrapped 
on (transverse to) the $i^{th}$ complex plane.}, and in a T-dual picture  
 of D3 and D7$_3$-branes,  to obtain the spectra of the models
in subsequent sections. In Section \ref{models} we systematically explore
the gauge group structure of models with discrete Wilson lines (Subsection
\ref{smodels}), and focus
on  a particular solution (Subsection \ref{semi}) with potentially
phenomenologically interesting features. We also present a solution
with anti-branes (Subsection \ref{anti}), which provides a  
realistic gauge group with three families.
 In Section \ref{conclusions} we  summarize and discuss  the 
results and techniques developed in this paper,  and  conclude with 
possible further applications. 

\section{Wilson lines in the $\IZ_6$ orientifold}
\label{wl}

The techniques to construct $N=1$ Type IIB orientifolds based on orbifolds
$T^6/\IZ_N$ and $T^6/(\IZ_N\times \IZ_M)$ are by now standard and we refer
the reader to, e.g., \cite{afiv} for a detailed discussion of such
models as well as the notations and conventions.

We shall focus on the study of Wilson lines for $\IZ_6$ orientifold, 
constructed \cite{kakuz6,afiv} by orientifolding 
 type IIB string theory on $T^6/\IZ_6$.  The $\IZ_6$ symmetry is generated 
by an action $\theta$: 
\beq 
  \theta: X_i \to e^{2i\p v_i}X_i \, , 
\eeq
on the complex coordinates $X_i$, $i=1,2,3$ of the three two-tori and 
the twist vector $v$ is of the form $v=\textstyle{1\over 6}(1,1,-2)$. The
orientifolding
is generated by the world-sheet parity operator $\Omega$. The model 
contains a set of D9- and D5$_3$-branes. 
The action  of $\theta$ and $\Omega$ on the open string states is
described by 
Chan-Paton matrices $\g_{\q}$  and  $\g_{\Omega}$, respectively,  which
form a projective
representation of the orientifold group.  Hence, 
consistency   with the group multiplication rules implies algebraic
constraints like
$\g_{\q}^N=\pm 1$, ($N=6$ 
for $\IZ_6$ orientifold), $(\g_{\q}^k)^*=\pm\g_{\O}^*\g_{\q}^k\g_{\O}$, etc. 
In addition,  $\g_{\O}$ must be symmetric for
D9-branes and antisymmetric for
D5-branes. In addition, Ramond-Ramond (RR)  sector tadpole cancellation
conditions must be imposed
on the Chan-Paton matrices to ensure the consistency of the theory. After 
successful embedding of the orientifold action, the open string states,
including gauge bosons and matter fields, can be easily constructed. 
 
We shall start with the analysis for the $\IZ_6$ orbifold, and only at a
later stage incorporate the orientifold projection $\Omega$. Also in this
section we shall focus primarily on the algebraic constraints on the
model, i.e. we discuss the building blocks for the Wilson lines, leaving
the
study of tadpole cancellation conditions for the subsequent Sections 
\footnote{This approach has an additional advantage that the results in
this Section 
apply to more general sets of branes, e.g., D2-branes wrapped on a complex 
plane, which could be of interest to study the spectrum of states in such
models.}.

The action of $\theta$ on D9-branes, $\g_{\theta,9}$, has the general form
\beqa
\gamma_{\theta,9}&=&\diag(e^{\pi i\frac 16}\id_{N_1}, e^{\pi i \frac 36}
\id_{N_2}, e^{\pi i \frac 56} \id_{N_3}, e^{\pi i \frac 76}\id_{N_4},
e^{\pi i \frac 96} \id_{N_5}, e^{\pi i \frac {11}{6}} \id_{N_6}) \, .
\label{CPtheta}
\eeqa
For the time being the non-negative integers $N_i$ are kept arbitrary, but
will be eventually constrained by the orientifold projection and tadpole 
cancellation conditions.

In the following we shall construct different types of discrete Wilson
lines on the D9-branes. Since the third complex plane is twisted only by
order three actions, the corresponding Wilson lines are identical to those
in the $\IZ_3$ orientifold \cite{ibanez,afiv,cl}. Hence we focus on
Wilson lines along the first complex plane (the second plane suffers
identical twists). The (space) group law implies the following set of
algebraic constraints on the Chan-Paton embedding $\gamma_{W,9}$
such Wilson lines:
\beqa
(\gamma_{\theta,9}\gamma_{W,9})^6 & = & -\id \, ;\nonumber\\
(\gamma_{\theta^2,9}\gamma_{W,9})^3 & = & -\id \, ; \nonumber \\
(\gamma_{\theta^3,9}\gamma_{W,9})^2 & = & -\id \, . 
\label{constraint}
\eeqa 
In the following subsections we construct several solutions to these
constraint, including order three Wilson lines,  order two
Wilson lines, and continuous Wilson lines. 

\subsection{\bf Order Three Wilson lines}
\label{wlo3}

An order three Wilson line $\gamma_{W_1,9}$ is characterized by
 the additional condition
\beq
\left[\gamma_{\theta^2,9}, \gamma_{W_1,9}\right]=0 \, , 
\label{commu1}
\eeq
hence the second constraint in (\ref{constraint}) implies
$\g_{W_1,9}^3=\id$. Given the relation between Wilson line eigenvalues and
brane locations in a T-dual picture, in the model obtained by T-dualizing
along the first complex plane, the configuration corresponds to locating
the D7-branes at different fixed points of the order three action
$\theta^2$. 

In the following we construct a family of Wilson lines satisfying these
constraints. It is convenient to reorder the blocks in
$\gamma_{\theta,9}$, so that we have
\beqa
\gamma_{\theta,9} &=&\diag(e^{\pi i\frac 16}\id_{N_1}, 
e^{\pi i \frac 76}\id_{N_4}, e^{\pi i \frac 36} \id_{N_2}, 
e^{\pi i \frac 96} \id_{N_5}, e^{\pi i \frac 56} \id_{N_3}, 
e^{\pi i \frac {11}{6}} \id_{N_6})  \, ; \nonumber \\
\gamma_{\theta^2,9} & = & \diag(e^{\pi i\frac 13}\id_{N_1}, 
e^{\pi i \frac 13}\id_{N_4}, - \id_{N_2}, -\id_{N_5},
e^{\pi i \frac 53} \id_{N_3}, e^{\pi i \frac 53} \id_{N_6}) \, ; \nonumber \\
\gamma_{\theta^3,9} & = & \diag(i \id_{N_1}, -i\id_{N_4}, -i\id_{N_2},
i\id_{N_5}, i\id_{N_3}, -i \id_{N_6}).
\label{gamma2}
\eeqa
The most general form of $\g_{W_1,9}$ which satisfies the condition 
(\ref{commu1}) is 
\beqa
\gamma_{W_1,9}  =  {\pmatrix{
\gamma_{W_1,9}^A & & \cr
& \gamma_{W_1,9}^B & \cr
& & \gamma_{W_1,9}^C \cr}} \, , 
\label{Chanpatons}
\eeqa
where each piece in the matrix above spans two blocks in (\ref{gamma2}).
The second constraint in Eqn.(\ref{constraint}) becomes
$(\g_{W_1,9}^{(p)})^3=\id$, $(p)=A,B,C$. One can show that the third
constraint in (\ref{constraint}) implies the first, so we need to impose
the former only.

The matrices $\gamma_{W_1,9}^{(p)}$, $p=A,B,C$ may contain diagonal blocks
of identity matrices with various dimensionalities, which correspond to
entries for which the Wilson line has a trivial action. In the following
discussion, we will tacitly remove  those entries, and define $N_i^{'}$ to
number the entries with  non-diagonal block structures.

There are non-trivial solutions for the constraints above when the 
(redefined) $N_i^{'}$ satisfy  $N_1^{'}=N_4^{'}$, $N_2^{'}=N_5^{'}$,
$N_3^{'}=N_6^{'}$. For instance, let us consider 
\beqa
& \gamma_{W_1,9}^A = {\pmatrix{
\frac 12 (A+A^2) & \frac 12 (A-A^2) \cr
\frac 12 (A-A^2) & \frac 12 (A+A^2) \cr
}} \quad ; \quad
\gamma_{W_1,9}^B = {\pmatrix{
\frac 12 (B+B^2) & \frac 12 (B-B^2) \cr
\frac 12 (B-B^2) & \frac 12 (B+B^2) \cr
}} ;  & \nonumber \\
& & \nonumber\\
& \gamma_{W_1,9}^C = {\pmatrix{
\frac 12 (C+C^2) & \frac 12 (C-C^2) \cr
\frac 12 (C-C^2) & \frac 12 (C+C^2) \cr
}} \, , &
\label{orderthree}
\eeqa
where $A$, $B$ and $C$ are diagonal matrices satisfying $A^3=\id$,
$B^3=\id$, $C^3=\id$. A direct  calculation, starting from constraints
in Eqn.(\ref{constraint}) and 
Eqn.(\ref{commu1}),  confirms that the building blocks
$\gamma_{W_1,9}^{A,B,C}$  of the 
the Wilson line   (\ref{Chanpatons}),  are indeed of the form
given  in Eqn.(\ref{orderthree}), and
therefore define a consistent order three Wilson line for  the
$\IZ_6$ orbifold.

To clarify the structure of our solution, let us diagonalize $A$, $B$ and
$C$ matrices,  while maintaining  $\g_{\theta^2,9}$ diagonal, but making
$\gamma_{\theta^3,9}$ non-diagonal. Denoting matrices in the new basis
with a tilde, we have
\beqa
{\tilde A}=\diag(\id_{n_1},\alpha \id_{m_1},\alpha^2 \id_{m_1}) \;\; ;
\;\;
{\tilde B}=\diag(\id_{n_2},\alpha \id_{m_2},\alpha^2 \id_{m_2}) \;\; ;
\;\;
{\tilde C}=\diag(\id_{n_3},\alpha \id_{m_3},\alpha^2 \id_{m_3}) \, , 
\label{abc}
\eeqa
with $\alpha=e^{2\pi i/3}$. 
Hence we have
\beq
{\tilde \g}_{W_1, 9} = \diag(\id_{N_1+N_4-2m_1}, \alpha \id_{m_1},\alpha^2
\id_{m_1}, 
\id_{N_2+N_5-2m_2}, \alpha \id_{m_2},\alpha^2 \id_{m_2}, \id_{N_3+N_6-2m_3}, 
\alpha \id_{m_3},\alpha^2 \id_{m_3}),
\label{w1}
\eeq
where $N_{i}$ are as originally defined in Eqn.(\ref{gamma2}),
while $m_1$, $m_2$ 
and $m_2$ are non-negative integers satisfying $N_1+N_4-2m_1 \ge 0$,
$N_2+N_5-2m_2\ge 0$ and $N_3+N_6-2m_3\ge 0$. On the other hand 
\beq
{\tilde \gamma}_{\q^3,9}  =  i\ \diag(T_{11}, T_{22}, T_{33}), 
\eeq
where the matrix $T_{11}$ is defined as 
\beq
T_{11}  =  \left(
	\begin{array}{cccc}
\id_{N_1-m_1} & & &   \\
 & -\id_{N_4-m_1} & &   \\
       & & & \id_{m_1}  \\
       & & \id_{m_1} &  \\
\end{array} \right),  
\eeq
and $T_{22}$ and $T_{33}$ are defined analogously. 

In this basis the structure of the Wilson line 
nicely reflects the configuration of D7-branes in the T-dual picture. Each
eigenspace of ${\tilde \g}_{W_1,9}$ corresponds to a set of D7-branes at a
$\theta^2$ fixed points. The action of ${\tilde \g}_{\theta^3}$ leaves
invariant the eigenspace with ${\tilde \g}_{W_1}$-eigenvalue $1$, and
swaps the eigenspaces of ${\tilde \g}_{W_1}$-eigenvalues $\alpha$,
$\alpha^2$. In the T-dual picture, this corresponds to the fact that
$\theta^3$ action 
leaves  D7-branes at the origin  fixed, but it  exchanges D7-branes at the
remaining two $\theta^2$ fixed points.

\medskip

We conclude this subsection by studying the introduction of a second discrete
order three Wilson line $\g_{W_2, 9}$ in the second complex plane. This
Wilson line has to satisfy all the conditions in Eqs.(\ref{constraint}) and
Eqn.(\ref{commu1}), and in addition has to commute with the first Wilson line 
action, i.e. $\left[\gamma_{W_2,9}, \gamma_{W_1,9}\right]=0$. In the
basis in which $\gamma_{W_1,9}$ is diagonal 
these constraints can be easily
satisfied by the following form of $\gamma_{W_2,9}$ (for the sake of simplicity we
have omitted the  notation with a tilde):  
\beq
\gamma_{W_2,9} = \diag(Q_{11}, Q_{22}, Q_{33}), 
\label{w2}
\eeq
where the matrix $Q_{11}$ is defined as
\beq
Q_{11} = \diag(q_{11}, q_{12}, q_{13}), 
\eeq
the matrices $q_{1i}$ ($i=1,2,3$) are defined as 
\beq
q_{11}= \left( 
	\begin{array}{ccc}
	\id_{N_1+N_4-2m_1-2x_1} & & \\
	& \frac 12 (D+D^2) & \frac 12 (D-D^2) \\
	& \frac 12 (D-D^2) & \frac 12 (D+D^2) 
	\end{array}
\right); 
\eeq
\beq
q_{12}= \diag(\id_{m_1-x_2-x_3}, \alpha \id_{x_2}, \alpha^2 \id_{x_3}) \quad ;
\quad q_{13}= \diag(\id_{m_1-x_2-x_3}, \alpha^2 \id_{x_2}, \alpha \id_{x_3}), 
\eeq
where $D^3=1$ and $D$ is a matrix of dimension $x_1\times x_1$, while $x_2
+ x_3 \leq m_1$.
Matrices $Q_{22}$ and $Q_{33}$ can be defined analogously with a set of
integer parameters $x_{i}$ ($i=1,\ldots,12$).  

\subsection{\bf Order two Wilson lines}
\label{wlo2}

We now turn to the generic structure of order two Wilson lines. 
An order two Wilson line in the first complex plane acts on the D9-branes
with Chan-Paton matrix $\gamma_{W2_1,9}$. This embedding must satisfy the
set of algebraic constraints in Eqn.(\ref{constraint}). In addition, it
must satisfy the commutation relation
\beq
\left[\gamma_{\theta,9}, \gamma_{W2_1,9}\right]=0\; . 
\label{commu2}
\eeq
Thus,  the third constraint in (\ref{constraint}) becomes
$\g_{W2_1,9}^2=\id$. These Wilson lines correspond in a T-dual picture
to  branes sitting at the $\theta^3$ fixed points.

It is convenient to reorder the Chan-Paton blocks as follows
\beqa
\gamma_{\theta,9} &=&\diag(e^{\pi i\frac 96}\id_{N_5}, 
e^{\pi i \frac 16}\id_{N_1}, e^{\pi i \frac 56} \id_{N_3}, 
e^{\pi i \frac 36} \id_{N_2}, e^{\pi i \frac {11}6} \id_{N_6}, 
e^{\pi i \frac 76} \id_{N_4})  \; ;\nonumber \\
\gamma_{\theta^2,9} & = & \diag(e^{\pi i}\id_{N_5}, 
e^{\pi i \frac 13}\id_{N_1}, e^{\pi i \frac 53} \id_{N_3},
e^{\pi i}\id_{N_2}, e^{\pi i \frac 53} \id_{N_6}, e^{\pi i \frac 13}
\id_{N_4}) \; ;\nonumber \\
\gamma_{\theta^3,9} & = & \diag(i \id_{N_5}, i\id_{N_1}, i\id_{N_3}, 
-i\id_{N_2}, -i\id_{N_6}, -i \id_{N_4})\; .
\label{gamma3}
\eeqa
The general form of a Wilson line matrix, satisfying Eqn.(\ref{commu2}), 
is of the form
\beqa
\gamma_{W2_1,9} & = & {\pmatrix{
\g_{W2_1,9}^A & \cr
& \g_{W2_1,9}^B \cr }} \, ,
\eeqa
where each piece in the above matrix spans three blocks in  
(\ref{gamma3}).
The third constraint in (\ref{constraint}) becomes $(\g_{W2_1,9}^{(p)})^2 
=\id$, for $(p)=A,B$.

The constraints can be solved independently for the two building blocks
$\g_{W2_1}^{A}$ and  $\g_{W2_1}^{B}$. Since they are analogous, we detail
only the `A' block. Let us denote $\gamma_{\theta^k,9}^A$, the relevant
sub-block
in the $\IZ_6$ Chan-Paton embeddings. 

The matrices $\gamma_{W2_1,9}^{(p)}$, $p=A,B$ may contain diagonal pieces.
They would correspond to entries for which the Wilson line has a trivial
action. We again redefine the $N_i$ to number the entries with
non-diagonal block structures. 

There are non-trivial solutions for the constraints above when the
(redefined) $N_i$'s satisfy  $N_1=N_3=N_5$, $N_2=N_4=N_6$. In analogy
with the order three Wilson lines, it is convenient to go to  the basis 
in which  $\gamma_{W2_1,9}$ and $\gamma_{\theta^3,9}$ are  
simultaneously diagonalized, while other matrices like $\gamma_{\theta,9}$
become off-diagonal. We denote the matrices in this new basis by a
tilde. The structure of ${\tilde \gamma}^{A}_{W2_1,9}$ and ${\tilde
\gamma}^A_{\theta^3,9}$ is
\beqa
{\tilde \gamma}^A_{\theta^3,9} & = & \diag ( i\id,i\id,i\id) \; ; \nonumber \\
{\tilde \gamma}^A_{W_1,9} & = & \diag(A_1,A_2,A_3)\; , 
\eeqa
where $A_1$, $A_2$, $A_3$ are diagonal and square to $\id$.

In this basis the (non-diagonal) matrix ${\tilde \gamma}_{\theta^2,9}^A$
has the structure
\beqa
{\tilde \gamma}^A_{\theta^2,9}=-{\pmatrix{
  & \id & \cr  &     & \id \cr \id & & \cr }}\; .
\label{permut}
\eeqa
In order to impose the second constraint in (\ref{constraint}), we obtain
\beqa
({\tilde \gamma}^A_{\theta^2,9} {\tilde \gamma}^A_{W_1,9})^3=-{\pmatrix{
A_2 A_3 A_1 & & \cr & A_3 A_1 A_2 & \cr & & A_1 A_2 A_3 \cr }}\; .
\eeqa
Thus,  we get the constraint $A_1 A_2 A_3=\id$ (since the matrices $A_i$
are
diagonal and hence commute, one  automatically obtains that  $A_2 A_3
A_1=\id$ and $A_3 A_1 A_2=\id$ as well.).

The matrix ${\tilde \gamma}^A_{\theta,9}$ can be obtained as $-{\tilde
\gamma}^A_{\theta^3,9} ({\tilde \gamma}^A_{\theta^2,9})^2$, hence 
 the first constraint in (\ref{constraint}) is automatically satisfied.

In order to express these matrices in the original basis, where all the
$\gamma_{\theta^k,9}$ are diagonal, and $\gamma_{W2_1,9}$ is non-diagonal,
we change the basis using the unitary transformation
\beqa
P=\frac 1{\sqrt 3} \pmatrix{ 1 & 1 & 1 \cr 1 & \alpha & \alpha^2 \cr 1 &
\alpha^2 
& \alpha}\; .
\eeqa
We get
\beqa
P {\tilde \gamma}^A_{\theta^2,9} P^{-1}=-\diag(\id, \alpha^2 \id, \alpha
\id)\; ,
\eeqa
which is precisely the form  for $\gamma_{\theta^2,9}$ in the
initial basis. In the initial basis the Wilson line has the following
structure
\beqa
\gamma^A_{W2_1,9}=P{\tilde \gamma}^A_{W2_1,9} P^{-1}={\pmatrix{
A & A^{\alpha^2} & A^{\alpha} \cr A^{\alpha} & A & A^{\alpha^2} \cr
A^{\alpha^2} & A^{\alpha} & A \cr 
}}\; ,
\label{o2dw}
\eeqa
where we have defined
\beqa
A=\frac 13(A_1+A_2+A_3) \quad ; \quad A^{\alpha}= \frac 13 (A_1+\alpha
A_2+\alpha^2 A_3) \quad ; \quad A^{\alpha^2}= \frac 13 (A_1+\alpha^2 A_2 +
\alpha A_3)\; .
\eeqa
Naturally,  this Wilson line  satisfies all the
algebraic constraints (\ref{constraint}).

The structure of these Wilson lines (particularly in the rotated
basis, where these Wilson lines are diagonal) 
reflects the distribution of D7-branes in the T-dual picture. Since the
Wilson line commutes with the $\theta^3$ twist, in the T-dual
picture D7-branes are
at $\theta^3$ fixed points. The order three permutation action
(\ref{permut}) corresponds, in the T-dual picture, to the 
permutation action of $\theta^2$ on the three $\theta^3$ fixed points away
from the origin.

\subsection{\bf Continuous Wilson line in a $\IZ_6$ plane}
\label{cwl}

In this section we compute the 
structure of a continuous Wilson line embedded along the first plane,
$\g_{Wc,9}$. It is convenient to reorder the blocks in the Chan-Paton
matrices as
\beqa
\gamma_{\theta,9} & = & \diag(e^{\pi i\frac 16}\id_{N_1},
e^{\pi i \frac 96} \id_{N_5}, e^{\pi i \frac 56} \id_{N_3},
e^{\pi i \frac 76}\id_{N_4}, e^{\pi i \frac 36} \id_{N_2},
e^{\pi i \frac {11}{6}} \id_{N_6})  \; ; \nonumber \\
\gamma_{\theta^2,9} & = & - \diag(e^{2\pi i\frac 23}\id_{N_1},
\id_{N_5}, e^{2\pi i \frac 13} \id_{N_3}, e^{2\pi i \frac 23}\id_{N_4}, 
\id_{N_2}, e^{2\pi i \frac 13} \id_{N_6}) \; ;  \nonumber \\
\gamma_{\theta^3,9} & = & \diag(i \id_{N_1}, 
i\id_{N_5}, i\id_{N_3}, -i\id_{N_4}, -i\id_{N_2}, -i \id_{N_6}) \; .
\label{gamma4}
\eeqa
The continuous Wilson line will involve an equal number of entries from
each diagonal block (i.e. a combination of D-branes with  the traceless embedding). 
Therefore, in the following discussion of the continuous Wilson line we
take all the $N_i$'s to be equal.

Since the structure of continuous Wilson lines, consistent with an order
three orbifold twist, is known \cite{ibanez,afiv,cl}, it is convenient to
use this information to solve the second constraint in (\ref{constraint}).
One then imposes the additional conditions to obtain the final form of the
Wilson line in the $\IZ_6$ case.

The order three twist can be written in the following form: 
\beqa
\gamma_{\theta^2} & = & -\diag(e^{2\pi i \frac 23}, 1, e^{2\pi i\frac 13},
e^{2\pi i\frac 23}, 1, e^{2\pi i\frac 13}) \otimes \id_{N}\; . 
\eeqa
The structure of a continuous Wilson line consistent with the order three
twist is
\beqa
{\tilde \gamma}_{W,9}= \diag(\gamma_{c,1};\gamma_{c,2}) \otimes \id_N\; ,
\label{ansatz}
\eeqa
where $\gamma_{c,1}$, $\gamma_{c,2}$ are two independent $3\times 3$ matrices, 
each of which depends on one complex parameter. They must satisfy
$(\g_{c,i} \g_{Z_3})^3=\id$, where $\g_{Z_3}=\diag(1,e^{2\pi i\frac 13},
e^{2\pi i\frac 23})$, and are of the general form, parameterized by one 
complex parameter, as given in \cite{cl}. (The one real parameter
continuous 
$\IZ_3$ Wilson line was first given in \cite{ibanez,afiv}.) We have
introduced  a tilde notation in the Wilson line above because it has such
a simple form only in a rotated basis, where the matrices $\gamma_{\theta}$, 
$\gamma_{\theta^3}$ become non-diagonal.
Specifically, the $\theta^3$ twist is embedded as
\beqa
{\tilde \gamma}_{\theta^3,9} & = & i\ {\pmatrix{  & \id_3 \cr \id_3 & }}
\otimes \id_N .
\eeqa
Hence, the second condition in (\ref{constraint}) requires
$\gamma_{c,2}=\gamma_{c,1}^{-1}$. From now on we denote $\gamma_{c,1}$,
$\gamma_{c,2}$ simply by $\gamma$, $\gamma^{-1}$, and also define
$\gamma_+\equiv\frac 12(\gamma+\gamma^{-1})$, $\gamma_-\equiv \frac
12(\gamma-\gamma^{-1})$.

Rotating back to the original basis, we have
\beqa
\gamma_{\theta,9} & = & \diag(e^{\pi i\frac 16}, e^{\pi i \frac 96}, 
e^{\pi i \frac 56}, e^{\pi i \frac 76}, e^{\pi i \frac 36}, e^{\pi i \frac
{11}{6}})  \; ; \nonumber \\
\gamma_{\theta^2,9} & = & - \diag(e^{2\pi i\frac 23}, 1, e^{2\pi i\frac
13}, e^{2\pi i \frac 23}, 1, e^{2\pi i \frac 13}) \; ; \nonumber \\
\gamma_{\theta^3,9} & = & \diag(i, i, i, -i, -i, -i) \; ; \nonumber \\
\gamma_{W,9} & = & {\pmatrix{\gamma_+ & \gamma_- \cr \gamma_- & \gamma_+ }} \; ,
\eeqa
in which only one diagonal block of the $\gamma_{\theta^k}$ matrices is
shown. The matrix $\gamma_{W,9}$ provides a one (complex) parameter Wilson
line solving the second and third constraints. A lengthy,  but
straightforward calculation shows that it also satisfies the first
constraint. Hence,
the matrix above provides a continuous Wilson line consistent with all
algebraic constraints. The Ansatz (\ref{ansatz}) can be extended to a more
general form, depending on $N$ complex parameters, as
\beqa
{\tilde \gamma}_{W,9}= \bigoplus_{i=1}^N
\diag(\gamma_{c_i};\gamma_{c_i}^{-1}) \, ,
\eeqa
with $\gamma_{c_i}$ a matrix of the form determined in \cite{cl}, and
depending on the $i^{th}$ complex parameter.

The construction above has a clear interpretation in the T-dual picture.
A continuous Wilson line corresponds to placing a dynamical D-brane at a
generic point on  the orbifold. Equivalently, such a D-brane corresponds
to a $\IZ_6$-invariant set  of six D-branes in the covering space. Imposing 
the order three constraint on the Wilson line amounts to grouping the T-dual 
D-branes in two $\IZ_3$-invariant sets. The additional order two condition
relates the locations of the two trios, yielding a $\IZ_6$-invariant
configuration.

At particular points in the moduli space of continuous Wilson lines, the
matrix $\gamma_{Wc,9}$ reproduces the structures encountered in the 
construction of order three and order two discrete Wilson lines. When the
relation $[\g_{\q^2}, \g_{W,9}]=0$ is imposed, one recovers the order
three discrete Wilson line: the matrix $\gamma$ can be diagonalized
simultaneously with $\g_{\q^2}$, hence it reduces to the previous 
 order three discrete Wilson line. In the case of order two discrete
Wilson lines, the relation $[\g_{\q^3}, \g_{W,9}]=0$ requires $\gamma = 
\gamma^{-1}$. The explicit form for the one complex parameter continuous 
Wilson line $\gamma$, as given in \cite{cl}, allows one to reduce $\gamma$
precisely to the form of the order two discrete Wilson line solution,
given in Eqn.(\ref{o2dw}), up to an irrelevant overall phase. 

In the T-dual picture, these special points in moduli space correspond to
configurations where the branes are located at fixed points of $\theta^2$
or $\theta^3$, as explained in the previous two Subsections.

\section{Tadpole Cancellation Conditions}
\label{tadpole}

Consistent  orientifold constructions also requires the conditions
of cancellation of RR  tadpoles. These have been
computed for the $\IZ_6$
orientifold in \cite{kakuz6,afiv} for the case without Wilson lines, and
require the introduction of D9- and D5$_3$-branes with constrained
Chan-Paton embeddings. Tadpole conditions can be  easily modified to
include the possibility of Wilson lines.
Crosscap contributions (twisted orientifold charges) are unchanged, since
the closed string sector does not feel the Wilson lines. Disk
contributions to twisted tadpoles at a fixed point are proportional to the
trace of the D-brane Chan-Paton embedding felt {\em locally} at the fixed
point, which depends on the presence of the Wilson lines.

Using the results in \cite{kakuz6,afiv} and the above properties 
of the one-loop diagrams in the presence of Wilson lines, it is
straightforward to write the tadpole cancellation conditions for 
general configurations in the presence of Wilson lines. For instance, we
consider configurations containing D9-branes, with Wilson lines along all
three complex planes,
and D5$_3$-branes, with Wilson lines along the third plane, and located at 
all possible $\theta^2$ fixed points in the first two planes (labelled
by $(m,n)$, with $m,n=0,\pm 1$ for shorthand). The twisted tadpole
conditions for the different fixed points are:
\beqa
{\bf {\rm Fixed\   point}} & ~~~ & {\bf {\rm Twisted\  Tadpole \ Condition}}\nonumber\\
(0,0,p)& & \Tr (\gamma_{\theta,9}\gamma_{W_3,9}^p)+\Tr
(\gamma_{\theta,5_3,(0,0)} \gamma_{W_3,5_3,(0,0)}^p) = 0 \; ; \nonumber \\
        & & \Tr (\gamma_{\theta^2,9} \gamma_{W_3,9}^{2p}) + 3 \Tr
(\gamma_{\theta^2,5_3,(0,0)} \gamma_{W_3,5_3,(0,0)}^{2p})= 16 \; ; \nonumber \\
       & & \Tr (\gamma_{\theta^3,9}) + 4 \Tr (\gamma_{\theta^3,5_3,(0,0)})
= 0 \; ; \nonumber \\
      & & \Tr (\gamma_{\theta^4,9} \gamma_{W_3,9}^{p}) + 3 \Tr
(\gamma_{\theta^4,5_3,(0,0)} \gamma_{W_3,5_3,(0,0)}^{p})= -16 \; ; \nonumber \\
      & & \Tr (\gamma_{\theta^5,9}\gamma_{W_3,9}^{2p})+\Tr
(\gamma_{\theta^5,5_3,(0,0)} \gamma_{W_3,5_3,(0,0)}^{2p}) = 0 \; ; \nonumber \\
\nonumber \\
(m,n,p) & & \Tr (\gamma_{\theta^2,9} \gamma_{W_1,9}^m \gamma_{W_2,9}^n
\gamma_{W_3,9}^{2p}) + 3 \Tr (\gamma_{\theta^2,5_3,(m,n)}
\gamma_{W_3,5_3,(m,n)}^{2p})= 4 \; ; \nonumber \\
{\scriptsize (m,n)\neq (0,0)} & & \Tr (\gamma_{\theta^4,9}
\gamma_{W_1,9}^{2m}
\gamma_{W_2,9}^{2n} \gamma_{W_3,9}^{p}) + 3 \Tr
(\gamma_{\theta^4,5_3,(m,n)} \gamma_{W_3,5_3,(m,n)}^{p})= -4 \; .
\label{tad1}
\eeqa

In subsequent sections we will provide a systematic search for solutions
to these conditions. In fact, it will be practical to consider the models
obtained after T-dualizing along the third complex plane, so that Wilson
lines along it become D-brane positions in the T-dual picture. The
original
D9-branes become D7$_3$-branes, sitting at the three $\theta^2$ fixed
points in the third plane, labelled by $p=0,\pm 1$, and the D5$_3$-branes
become D3-branes, sitting at $\theta^2$ fixed points $(m,n,p)$. The
orientifold action in the T-dual version is $\Omega_3=\Omega R_3
(-)^{F_L}$, with $R_3:z_3\to -z_3$. 

In order to construct the model in the T-dual version, we need to
derive the
corresponding tadpole cancellation conditions. These can be obtained
from the original ones, given in Eqn.(\ref{tad1}), by applying the Fourier
transformation procedure, discussed in the Appendix. Applying this
procedure along the third complex 
plane, the new tadpole conditions read:
\beqa
{\bf {\rm Fixed\  point}} &~~~& {\bf {\rm Twisted \ Tadpole \ Condition}}\nonumber\\
(0,0,p) &  & \Tr (\gamma_{\theta,7_3,p}) + \Tr
(\gamma_{\theta,3,(0,0,p)})=0 \; ; \nonumber \\
        & & \Tr (\gamma_{\theta^2,7_3,p}) + 3 \Tr
(\gamma_{\theta^2,3,(0,0,p)})= 16 \delta_{p,0} \; ; \nonumber \\
        & & \sum_{p=0,\pm 1}\, [\,\Tr (\gamma_{\theta^3,7_3,p}) + 
4 \Tr(\gamma_{\theta^3,3,(0,0,p)})\; ] = 0 \; ; \nonumber \\
        & & \Tr (\gamma_{\theta^4,7_3,p}) + 3 \Tr
(\gamma_{\theta^4,3,(0,0,p)})= -16 \delta_{p,0} \; ; \nonumber \\
        & & \Tr (\gamma_{\theta^5,7_3,p}) + \Tr
(\gamma_{\theta^5,3,(0,0,p)})=0 \; ; \nonumber\\
\nonumber \\
(m,n,p) & & \Tr (\gamma_{\theta^2,7_3,p} \gamma_{W_1,7_3,p}^m
\gamma_{W_2,7_3,p}^n) + 3 \Tr \gamma_{\theta^2,3,(m,n,p)} = 4 \delta_{p,0}
\; ; \nonumber \\
{\scriptsize (m,n)\neq(0,0)} & & \Tr (\gamma_{\theta^4,7_3,p}
\gamma_{W_1,7_3,p}^{2m}
\gamma_{W_2,7_3,p}^{2n}) + 3 \Tr \gamma_{\theta^2,3,(m,n,p)} 
= -4\delta_{p,0} \; .
\label{Ttadpole}
\eeqa

\begin{figure}
\begin{center} 
\centering
\epsfysize=8cm
\leavevmode
\epsfbox{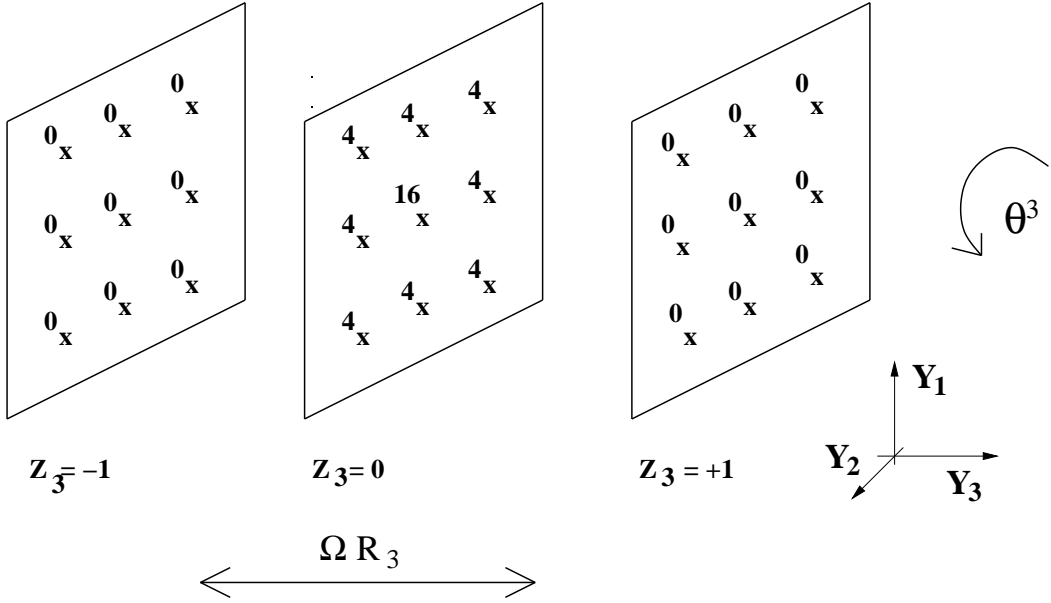}
\end{center}
\caption[]{\small Structure of $T^6/\IZ_6$ modded out by $\Omega_3$.
At each $\theta^2$ fixed point we have indicated the $\theta^2$-twisted
crosscap tadpole. This has to be canceled by including D3-branes (denoted
by crosses) and D7$_3$-branes (depicted as planes) sitting or passing
through those points.} 
\label{fig1}
\end{figure}

These conditions nicely reflect the geometric structure of the fixed
points in the T-dual picture, which is depicted in Fig. \ref{fig1}. It
represents the 27 $\theta^2$ fixed points of $T^6/\IZ_6$, each with its 
contribution to the $\theta^2$-twisted crosscap tadpole which, if
non-zero,
has to be canceled by D3 branes sitting at (or D7$_3$-branes passing
through) those points. Notice that crosscap tadpoles arise only at
fixed points of $\Omega_3$, which therefore must have $p=0$.
Recall also that $\theta^3$ reflects $z_1$ and $z_2$, so only points with
$m=n=0$ are fixed under it. Taking all these conditions into account,
we see the following properties of the tadpole cancellation conditions:

\begin{itemize}
\item The point $(0,0,0)$ is fixed under $\theta^2$, $\theta^3$ and
$\Omega_3$, it is a $\IC^3/\IZ_6$ orientifold point. The contributions to
the $\theta^k$ crosscap tadpoles for such point can be computed to be $0$, 
$16$, $0$, $-16$, $0$ for $k=1,...,5$, as above.
\item A point of the form $(m,n,0)$ with $(m,n)\neq (0,0)$ is fixed under
$\theta^2$ and $\Omega_3$, but not under $\theta^2$, and so is a $\IC^3/\IZ_3$
orientifold point. The crosscap tadpoles in its two twisted sectors are
$4$, $-4$ \cite{iru}, as above.
\item Points of the form $(0,0,p)$ for non-zero $p$, are fixed under
$\theta^2$, $\theta^3$, but not under $\Omega_3$; they are $\IC^3/\IZ_6$
orbifold points, and have zero crosscap tadpoles.
\item Points of the form $(m,n,p)$ for non-zero $p$ and $(m,n)\neq (0,0)$
are fixed under $\theta^2$ but not under $\theta^3$ or $\Omega_3$; they
are $\IC^3/\IZ_3$ orbifold points, and do not receive crosscap tadpoles. 
\end{itemize}

As discussed in \cite{Aldazabal}, $\IC^3/\IZ_3$ orbifold points are
interesting to build three-family models with realistic gauge groups on
the D3-branes. The $\IZ_3$ symmetry ensures three families structure, while the 
 orbifold (rather than orientifold) points ensures the existence of one
non-anomalous U(1) which could play the role of hypercharge, if the gauge groups arising from these D3-branes contain $SU(3)_{color}\times SU(2)_{L}\times U(1)$. The existence of $\IZ_3$ orbifold points 
is one of the motivations for studying $\IZ_6$ orientifold.

\section{Spectrum} 
\label{spectrum}

Before turning to the techniques of solving the tadpole
conditions and exploring their solutions, we devote this section to 
deriving the spectrum in the open string sectors by imposing   
orbifold and orientifold projections. 

The class of models we are considering, as mentioned above, includes
models with D9-branes with arbitrary order three Wilson lines along the
three complex planes, and D5$_3$-branes at the diverse $\theta^2$ fixed
points in the first two complex planes, and with an arbitrary order
three Wilson line on the third plane. 

In the 99 sector the projections for gauge bosons and chiral multiplets
are
\beqa
\begin{array}{cccc}
{\bf 99}\ & {\rm Gauge}\ & \lambda = \gamma_{\theta,9} \lambda
\gamma_{\theta,9}^{-1}\; ;\ & \lambda = - \gamma_{\Omega,9} \lambda^T
\gamma_{\Omega,9}^{-1} \; ; \\
         & i^{th} {\rm Chiral}\ & \lambda = e^{2\pi i v_i} 
\gamma_{\theta,9} \lambda \gamma_{\theta,9}^{-1}\; ;\ & \lambda =
- \gamma_{\Omega,9} \lambda^T \gamma_{\Omega,9}^{-1}\; , 
\end{array}
\eeqa
where $i=1,2,3$ labels the complex planes. Also, since
boundary conditions allow for momentum excitations in all complex planes,
we
have to impose the projection onto states invariant under the action
of the three Wilson lines
\beqa
{\bf 99}\ \ \lambda = \gamma_{W_i,9} \lambda \gamma_{W_i,9}^{-1} \ \ {\rm
for} \ {\rm all} \ 99\ {\rm states}\; ,
\label{projwl}
\eeqa
for $i=1,2,3$. In the $5_3 5_3$ sector, we obtain states only for open
strings stretching between the D5$_3$-branes sitting at identical points
$(m,n)$ in the first two planes. The projections for gauge bosons and
chiral multiplets are
\beqa
\begin{array}{cccc}
{\bf 5_3 5_3}\ & {\rm Gauge} & \lambda = \gamma_{\theta,5_3,(m,n)} \lambda
\gamma_{\theta,5_3,(m,n)}^{-1} \; ; & \lambda = - \gamma_{\Omega,5_3,(m,n)}
\lambda^T \gamma_{\Omega,5_3,(m,n)}^{-1} \; ; \\
         & {\rm Chiral} & \lambda = e^{2\pi i v_i} 
\gamma_{\theta,5_3,(m,n)} \lambda \gamma_{\theta,5_3,(m,n)}^{-1}\; ; & \lambda
= \pm \gamma_{\Omega,5_3,(m,n)} \lambda^T \gamma_{\Omega,5_3,(m,n)}^{-1}\; ,  
\end{array}
\eeqa
with $i=1,2,3$, and with the sign in the last equation positive for
$i=1,2$, and negative for $i=3$. Since boundary conditions allow
for momentum
excitations in the third complex planes, all states must satisfy
\beqa
{\bf 5_3 5_3}\ \ \lambda = \gamma_{W_3,5_3,(m,n)} \lambda
\gamma_{W_3,5_3,(m,n)}^{-1} \ \ {\rm for} \ {\rm  all} \ 5_35_3\ {\rm
states}\; . 
\eeqa

Finally, we turn to the 95$_3$ sector. There is one such sector for each
point $(m,n)$ at which D5$_3$-branes sit. Since $\Omega$ maps it to the
5$_3$9 sector, we keep only the former and do not impose any $\Omega$
projection. In performing the orbifold projection one should realize that
at a point with coordinates $(m,n)$ in the first two planes the $\theta$
action felt by the D9-branes depends on the Wilson lines along these
directions. We have the projection
\beqa
{\bf 95_3}\ \ & \lambda = e^{-\pi i v_3} ( \gamma_{\theta,9}
\gamma_{W_1,9}^m
\gamma_{W_2,9}^n) \lambda \gamma_{\theta,5_3,(m,n)}^{-1}\; .
\eeqa
In addition since boundary conditions allow for  momentum along the third
plane,
we
have to impose
\beqa
{\bf 95_3}\ \ \lambda = \gamma_{W_3,9} \lambda \gamma_{W_3,5_3,(m,n)}^{-1}
\ \ {\rm for} \ 95_3\ {\rm states}\; .
\eeqa

\medskip

In the  T-dual version, in which the
orientifold action is $\Omega_3$, and the model contains D7$_3$- and
D3-branes, it is convenient to describe the projections in these terms
as well. 
Our class of models contains D7$_3$-branes sitting at $\theta^2$ fixed
points in the third plane, and with arbitrary order three Wilson lines
along the first two, and D3-branes sitting at $\theta^2$ fixed points
$(m,n,p)$ in the internal space.

In the 7$_3$7$_3$ sector there are massless states only when both 7$_3$
branes sit at the same location in the third plane (this is T-dual to the 
$W_3$ projection in (\ref{projwl})). The only D7$_3$-branes that are left
fixed by $\Omega_3$ projection sit at $p=0$. For these, the projections
read
\beqa
\begin{array}{cccc}
{\bf 7_37_3}\ & {\rm Gauge}\ & \lambda = \gamma_{\theta,7_3,0} \lambda
\gamma_{\theta,7_3,0}^{-1}\; ; \ & \lambda = - \gamma_{\Omega_3,7_3,0}
\lambda^T \gamma_{\Omega_3,7_3,0}^{-1} \; ; \\
         & i^{th} {\rm Chiral}\ & \lambda = e^{2\pi i v_i} 
\gamma_{\theta,7_3,0} \lambda \gamma_{\theta,7_3,0}^{-1}\; ; \ & \lambda =
- \gamma_{\Omega_3,7_3,0} \lambda^T \gamma_{\Omega_3,7_3,0}^{-1}\; .  
\end{array}
\eeqa
For D7$_3$-branes at a location $p\neq 0$, $\Omega_3$ maps them to
D7$_3$-branes at the location $-p$. Hence, we may keep only one set of
them, say $p=+1$, and do not impose the orientifold projection. We have
\beqa
\begin{array}{ccc}
{\bf 7_37_3}\ & {\rm Gauge}\ & \lambda = \gamma_{\theta,7_3,+1} \lambda
\gamma_{\theta,7_3,+1}^{-1} \; ; \\
         & i^{th} {\rm Chiral}\ & \lambda = e^{2\pi i v_i} 
\gamma_{\theta,7_3,+1} \lambda \gamma_{\theta,7_3,+1}^{-1} \; . 
\end{array}
\eeqa

Since boundary conditions allow for momentum excitations in the first two 
complex planes, we have the Wilson line projection
\beqa
{\bf 7_37_3}\ \
\lambda = \gamma_{W_i,7_3,p} \lambda \gamma_{W_i,7_3,p}^{-1} \ \ {\rm for}
\ {\rm  all} \ 7_3 7_3\  {\rm states}
\eeqa
for $p=0,\pm 1$ and $i=1,2$.

In the 33 sector, we obtain states only for D3-branes sitting at
the identical
points $(m,n,p)$. Notice that all D3-branes sit at $\theta^2$ fixed
points, hence all 33 states are subject to  the $\theta^2$
projection.
However, D3-branes with $(m,n)\neq(0,0)$ are not fixed under $\theta$,
which maps them to D3-branes at $(-m,-n)$, so we may keep only the
$\theta^2$ projection 
and do not impose the $\theta$ projection. Similarly, only D3-branes with
the third coordinate $p=0$ are fixed under $\Omega_3$, and thus are
subject to $\Omega_3$ projection.  In the following we state the explicit
projections for D3-branes at different $\theta^2$ fixed points.

For D3-branes at $(0,0,0)$  point we have the projections
\beqa
\begin{array}{cccc}
{\bf 33} & {\rm Gauge} & \lambda = \gamma_{\theta,3,(0,0,0)} \lambda
\gamma_{\theta,3,(0,0,0)}^{-1} \; ; & \lambda = - \gamma_{\Omega_3,3,(0,0,0)}
\lambda^T \gamma_{\Omega_3,3,(0,0,0)}^{-1} \; ;  \\
         & i^{th} {\rm Chiral} & \lambda = e^{2\pi i v_i} 
\gamma_{\theta,3,(0,0,0)} \lambda \gamma_{\theta,3,(0,0,0)}^{-1} \; ; & \lambda
= \pm \gamma_{\Omega_3,3,(0,0,0)} \lambda^T \gamma_{\Omega_3,3,(0,0,0)}^{-1} \; ,  
\end{array}
\eeqa
with positive sign for $i=1,2$, and negative for $i=3$. 

For D3-branes at $(m,n,0)$ with $(m,n)=(1,0), (0,1), (1,1), (1,-1)$, we
have the projections
\beqa
\begin{array}{cccc}
{\bf 33}\ & {\rm Gauge} & \lambda = \gamma_{\theta^2,3,(m,n,0)}\ \lambda\
\gamma_{\theta^2,3,(m,n,0)}^{-1} \; ; & \lambda = - \gamma_{\Omega,3,(m,n,0)}\
\lambda^T\ \gamma_{\Omega,3,(m,n,0)}^{-1} \; ;  \\
         & i^{th} {\rm Chiral} & \lambda = e^{2\pi i 2v_i}\ 
\gamma_{\theta^2,3,(m,n,0)}\ \lambda\ \gamma_{\theta^2,3,(m,n,0)}^{-1} \; ; &
\lambda = \pm \ \gamma_{\Omega,3,(0,0,0)}\ \lambda^T\
\gamma_{\Omega,3,(0,0,0)}^{-1} \; , 
\end{array}
\eeqa
with positive sign for $i=1,2$, and negative for $i=3$. Other 
%non-zero
$(m,n)$  fixed points are just $\theta$-images of the above.

For D3-branes at $(0,0,+1)$ 
\beqa
\begin{array}{ccc}
{\bf 33}\ & {\rm Gauge} & \lambda = \gamma_{\theta,3,(0,0,+1)}\ \lambda\
\gamma_{\theta,3,(0,0,+1)}^{-1} \; ; \\
         & i^{th} {\rm Chiral} &\ \ \lambda = e^{2\pi i v_i}\ 
\gamma_{\theta,3,(0,0,+1)}\ \lambda\ \gamma_{\theta,3,(0,0,+1)}^{-1}\ 
\gamma_{\Omega,3,(0,0,0)}^{-1} \; . 
\end{array}
\eeqa
D3-branes at $p=-1$ are just $\Omega_3$-images of the above.

Finally, for D3-branes at $(m,n,+1)=(0,1,1), (1,0,1), (1,1,1), (1,-1,1)$,
we have
\beqa
\begin{array}{cccc}
{\bf 33} & {\rm Gauge} & \lambda = \gamma_{\theta^2,3,(m,n,+1)}\ \lambda\
\gamma_{\theta^2,3,(m,n,+1)}^{-1} \; ; \\
         & i^{th} {\rm Chiral} &\ \lambda = e^{2\pi i 2v_i} \
\gamma_{\theta^2,3,(m,n,+1)}\ \lambda\ \gamma_{\theta^2,3,(m,n,+1)}^{-1} 
\; . 
\end{array}
\eeqa
Other   $(m,n,+1)$ points  are just $\theta$-images of the above, and 
points $(m,n,-1)$ are $\Omega_3$-images of the above.

Finally, we turn to the 7$_3$3 sector. There is one such sector for each
point $(m,n,p)$ at which D3-branes may sit. The  massless states arise
from D7$_3$- and D3-branes with identical $p$. In performing the orbifold
projection one should realize that at a point with coordinates $(m,n)$ in
the first two planes the $\theta$ action felt by the D7$_3$-branes depends
on the Wilson lines along these directions. Again, we have to
distinguish several cases.

For $(0,0,0)$, $\Omega_3$ maps the 7$_3$3 sector to the 37$_3$ sector, 
so we keep
the former and do not impose the $\Omega_3$ projection. We have
\beqa
{\bf 7_33}\ \  & \lambda = e^{-\pi i v_3}\ \gamma_{\theta,7_3,0}\
\lambda\ \gamma_{\theta,3,(0,0,0)}^{-1} \; .
\eeqa

At points $(m,n,0)$, $\Omega_3$ maps the 7$_3$3 sector to  the 37$_3$ one,
so
we keep the former and  do not impose the $\Omega_3$ projection. Also,
$\theta$ action
maps the point $(m,n,0)$ to $(-m,-n,0)$, and we may restrict ourselves  to
$(m,n)=
(1,0), (0,1), (1,1), (1,-1)$. The $\theta^2$ projection is
\beqa
{\bf 7_33}\ \  & \lambda = e^{-\pi i 2v_3}\ ( \gamma_{\theta^2,7_3,0}\
\gamma_{W_1,7_3,0}^{2m}\ \gamma_{W_2,7_3,0}^{2n})\  \lambda\ 
\gamma_{\theta,3,(m,n,0)}^{-1} \; .
\eeqa

At points $(0,0,p)$, $\Omega_3$ maps the 7$_3$3 sector to the 3$7_3$
sector {\em at a different point} $(0,0,-p)$, so we may keep both the
37$_3$ and 7$_3$3 sectors at $(0,0,1)$ and do not impose the $\Omega_3$
projection. We have
\beqa
{\bf 7_33}\ \  & \lambda = e^{-\pi i v_3}\ \gamma_{\theta,7_3,+1}\
\lambda\ \gamma_{\theta,3,(0,0,+1)}^{-1} \; ; \nonumber \\
{\bf 37_3}\ \  & \lambda = e^{-\pi i v_3}\ \gamma_{\theta,3,(0,0,+1)}\
\lambda\ \gamma_{\theta,7_3,0}^{-1} \; .
\eeqa

Finally, at points $(m,n,p)$, we may keep both the 37$_3$ and 
7$_3$3 sectors for points with $(m,n)=(1,0), (0,1), (1,1), (1,-1)$ and
$p=+1$, and do not impose the $\theta$ or $\Omega_3$ projections. We
have
\beqa
{\bf 7_33}\ \  & \lambda = e^{-\pi i 2v_3}\ ( \gamma_{\theta^2,7_3,+1}\
\gamma_{W_1,7_3,+1}^{2m}\ \gamma_{W_2,7_3,+1}^{2n})\  \lambda\
\gamma_{\theta,3,(m,n,+1)}^{-1} \; ; \nonumber \\
{\bf 37_3}\ \  & \lambda = e^{-\pi i 2v_3}\ \gamma_{\theta^2,3,(m,n,+1)}\
\lambda\ (\gamma_{\theta,7_3,+1}\ \gamma_{W_1,7_3,+1}^{2m}\
\gamma_{W_2,7_3,+1}^{2n})^{-1} \; .
\eeqa

\section{Models}
\label{models}

\subsection{Systematic Construction of Models}
\label{smodels}
In this Section, we develop an algorithm which allows us to systematically
construct a large class of $\IZ_6$ orientifold models that satisfy the
algebraic and tadpole cancellation constraints.

In the $\IZ_6$ orientifold model constructed in \cite{kakuz6,afiv} which
does not have Wilson lines, all the D5$_3$ branes are placed at the
origin. The resulting gauge structure is $(U(6)\times U(4) \times U(4))^2$
from both D9- and D5$_3$ brane sector. To further break down the gauge
structure to smaller groups, Wilson lines have to be introduced. 

Our models in general include non-trivial discrete Wilson lines along the
third complex plane, embedded both on D9- and D5$_3$-branes. The
configuration becomes more intuitive if we perform a T-duality along this
plane. In the T-dual picture, there are D7$_3$-branes sitting at
three different fixed points in the third plane, $p=-1, 0, 1$. And there are 
D3-branes that sit at fixed points of all three complex planes, $(m,n,p)$. 
The tadpole cancellation conditions in Eqn.(\ref{Ttadpole}) together with the 
group consistency conditions constraint the number of the branes at each
fixed point and the gauge groups that arise from the set of branes. 

Consider D7$_3$ branes at $p=0$. Since $p=0$ is a $\IC^3/\IZ_6$ 
orientifold point, the action of $\theta$ twist takes the form of
Eqn.(\ref{CPtheta}). In addition, we have to solve the tadpole conditions
(\ref{Ttadpole}). A simple possibility is to saturate the $\theta^2$
tadpoles at the points $(m,n,0)$ with $(m,n)\neq (0,0)$ purely by the
D7$_3$-brane contribution (so that no D3-branes are required there),
satisfying $\Tr (\g_{\q^2, 7_3, 0})=4$. We choose to put $8$ $D7_3$
branes at $p=0$, with
\beq
\g_{\q, 7_3, 0} = \diag(e^{i\frac{\pi}{6}} \id_{2}, e^{i\frac{5\pi}{6}}
\id_{2}, e^{i\frac{7\pi}{6}} \id_{2},  e^{i\frac{11\pi}{6}} \id_{2}).
\label{chorigin} 
\eeq
In analogy with the analysis in \cite{abiu}, there exist other choices
also satisfying the tadpole conditions, but we do not consider them here.

We now turn to tadpoles at $(0,0,0)$, which are not fully canceled by the
D7$_3$-branes above. To cancel the remaining contribution, we introduce a
set of $D3$ branes at $(0,0,0)$, with
\beq
\g_{\q, 3, (0, 0, 0)} = \diag(e^{i\frac{\pi}{6}} \id_{2},
e^{i\frac{5\pi}{6}} \id_{2}, e^{i\frac{7\pi}{6}} \id_{2},
e^{i\frac{11\pi}{6}} \id_{2}). 
\label{chodthree}
\eeq 
Out of the remaining $24$ $D7_3$ branes left, we locate half of them 
at $p=1$ and the other half at its orientifold image $p=-1$. The $\q$ twist 
takes the general form of Eqn.(\ref{CPtheta})
\beq
\gamma_{\theta,7_3, 1}=\diag(e^{\pi i\frac 16}\id_{N_1}, e^{\pi i \frac 36}
\id_{N_2}, e^{\pi i \frac 56} \id_{N_3}, e^{\pi i \frac 76}\id_{N_4},
e^{\pi i \frac 96} \id_{N_5}, e^{\pi i \frac {11}{6}} \id_{N_6}),    
\label{enes}
\eeq
where $\sum_{i=1}^{6}N_i =12$. 
Their contributions to the tadpole associated with fixed point $(0,0,1)$
must be canceled by that from $D3$ branes sitting at the fixed point, as
seen from Eqn.(\ref{Ttadpole}). The $\q$ action on this set of $D3$ branes
takes a similar form 
\beq
\gamma_{\theta,3,(0,0,1)}=\diag(e^{\pi i\frac 16}\id_{M_1}, e^{\pi i \frac 36}
\id_{M_2}, e^{\pi i \frac 56} \id_{M_3}, e^{\pi i \frac 76}\id_{M_4},
e^{\pi i \frac 96} \id_{M_5}, e^{\pi i \frac {11}{6}} \id_{M_6}),    
\label{emes}
\eeq  
where $M_{i}$ is a different set of integers and their sum, which is the
total number of $D3$ branes placed at $(0,0,1)$, is not fixed. 

The rest of the $D3$ branes $(24-2\sum_{i=1}^{6}M_i)$ can be placed at fixed 
points $(m, n, p)$, where $(m, n) \neq (0,0)$. We also consider $p \neq 0$
as well, since tadpoles at $(m,n,0)$ are already canceled. There are four
types of points: $(1,0,1)$ and its $3$ images [i.e. the $\theta^3$ image
$(-1,0,1)$, and their $\Omega_3$ images, $(\pm 1, 0,-1)$]; $(0,1,1)$ and
its $3$ images; $(1,1,1)$ and its $3$ images; and $(1,-1,1)$ and its $3$
images. These points are fixed under $\theta^2$ only, so we need to
specify its orbifold action, which takes the form:
\beqa
\g_{\q^2, 3, (1, 0, 1)} & = & -\diag(\id_{i_1},
e^{i\frac{4\pi}{3}}\id_{i_2}, e^{i\frac{2\pi}{3}}\id_{i_3}); \nonumber \\  
\g_{\q^2, 3, (0, 1, 1)} & = & -\diag(\id_{j_1},
e^{i\frac{4\pi}{3}}\id_{j_2}, e^{i\frac{2\pi}{3}}\id_{j_3}); \nonumber \\ 
\g_{\q^2, 3, (1, 1, 1)} & = & -\diag(\id_{k_1},
e^{i\frac{4\pi}{3}}\id_{k_2}, e^{i\frac{2\pi}{3}}\id_{k_3});\nonumber\\
\g_{\q^2, 3, (1, -1, 1)} & = & -\diag(\id_{l_1},
e^{i\frac{4\pi}{3}}\id_{l_2}, e^{i\frac{2\pi}{3}}\id_{l_3}),   
\label{multiplechoice}
\eeqa
where $\sum_{n=1}^{3}(2i_n+2j_n+2k_n+2l_n)=12-\sum_{i=1}^{6}M_i$.

In addition, we also consider the possibility of general order three discrete 
Wilson lines in the first and second complex planes. The Wilson line in the
first complex plane takes the form of Eqn.(\ref{w1}) with three parameters
$m_{1,2,3}$, and the Wilson line in the second complex plane takes the
form of Eqn.(\ref{w2}) with $12$ parameters $x_{i}$ ($i=1,\ldots,12$). 

We can now calculate the traces of the $\g$-matrices and express the tadpole 
cancellation conditions Eqs.(\ref{Ttadpole}) in terms of the set of
parameters $N_n$, $M_n$, $i_n$, $j_n$, $k_n$, $l_n$, $m_n$, $x_n$. Since the 
number of equations is less than the number of parameters, we search through  
the whole parameter space for allowed solutions. There is a seemingly large 
number of solutions, however, many solutions are related by symmetries of the 
construction, and are therefore equivalent. Our results show that the number 
of inequivalent solutions is limited. 

Let us briefly discuss some of these equivalences. A transformation
leaving the full spectrum of the theory invariant (including massive
states) is the following: Multiply the $\IZ_6$ Chan-Paton matrices
$\gamma_{\theta^k}$ of all D-branes at $z_3=+1$ by a constant
brane-independent phase $e^{i k2\pi n/6}$, where $n=1\dots 5$  
(and by the conjugated phase in
the orientifold images at $z_3=-1$), leaving the Wilson lines unchanged. 
Models related by this transformation are completely equivalent. A second
transformation which leaves the full theory invariant is to conjugate the
orbifold and Wilson line Chan-Paton matrices for all D-branes at $z_3=\pm
1$. Since matrices at $z_3=+1$ and their orientifold images are conjugated
of each other, the above operation is equivalent to the geometric symmetry
of reflecting the models with respect to the plane $z_3=0$. 

Imposing these equivalences, the number of solutions reduces drastically.
In particular, we find that among the whole set of solutions only a few
have a potential to give semi-realistic models, i.e. those that have a
gauge structure $SU(3)\times SU(2)^{i} \times U(1)^{j}$ as a subgroup. 

Interestingly enough, we found only a few inequivalent classes of 
semi-realistic  models containing a gauge structure of $U(3)\times U(2)^2
\times U(1)$ within a single D-brane sector (an interesting situation to
obtain family replication of the spectrum). It arises from a set of D3 branes 
located at the fixed point ($0,0,1$), or its T-dual version. The Chan-Paton 
  matrices for $\g_{\theta, 7_3, 1}$ and $\g_{\theta, 3, (0,0,1)}$
take the form
\begin{equation}
\begin{array}{ccc}
\g_{\theta, 7_3, 1}& = & \diag(e^{i\frac{\pi}{6}}\id_2,
e^{i\frac{3\pi}{6}}\id_3, e^{i\frac{5\pi}{6}}\id_4,
e^{i\frac{7\pi}{6}}\id_1, e^{i\frac{11\pi}{6}}\id_2) \; ; \\
\g_{\theta, 3,(0,0,1)}& = & \diag(e^{i\frac{\pi}{6}}\id_1,
e^{i\frac{7\pi}{6}}\id_2, e^{i\frac{11\pi}{6}}\id_3, 
e^{i\frac{11\pi}{6}}\id_2). 
\end{array}
\end{equation} 
The configuration and the Chan-Paton matrices of the remaining D3-branes and the
Wilson line actions in the model are listed in Table I. Using the
transformations above we can generate other models with seemingly
different Chan-Paton matrices, but with completely equivalent spectra.

Moreover, close inspection of Table I reveals that some of these solutions
are also related by further symmetries of the model, associated with
modular transformations in the four-torus associated with the first two
complex planes. For instance, the exchange of these complex planes has a
non-trivial action on fixed points [mapping $(m,n,p)$ to $(n,m,p)$] and on
Wilson lines [exchanging $\gamma_{W_1,7_3}$ and $\gamma_{W_2,7_3}$], and
allows one to relate, e.g., the third and fourth model, or the sixth
and
eighth model in Table I. Other modular transformations allow for less obvious
equivalences. For instance, there exists a transformation mapping the
fixed point $(m,n,p)$ to $(m+n,n,p)$, and acting on Wilson lines as
\beqa
\gamma_{W_1,7_3,1} \to \gamma_{W_1,7_3,1} \quad ; \quad
\gamma_{W_2,7_3,1} \to \gamma_{W_2,7_3,1} \gamma_{W_1,7_3,1}^{-1} \; .
\eeqa
This allows one to relate models like the second and third, or like the
fifth
and sixth in Table I. 
Employing all the modular transformation equivalences, we find there are
only two distinct solutions with the above mentioned gauge structure: 
the first contains two D3-branes placed at one fixed point $(m,n,1)$ with 
$(m,n)\neq (0,0)$, while the second involves two such fixed points, with
one D3-brane sitting at each point. They correspond to the two equivalence
classes of the models in Table I (which are separated by a double line).

In the following Subsection, we study in more detail the gauge structure
and
the matter spectrum of the first kind of the semi-realistic models, with two
D3-branes at one fixed point. The second model can be analyzed
analogously, and leads to very similar phenomenology.

\begin{table}
\begin{center}
\begin{tabular}{|c|c|c|c|c|c|}

($i_1, i_2, i_3$) & ($j_1, j_2, j_3$) & ($k_1, k_2, k_3$) & ($l_1, l_2, l_3$) &
($m_1, m_2, m_3$) & ($x_1, x_2, \dots, x_9$)  \\ \hline \hline

($0,0,0$) & ($0,0,0$) & ($0,0,0$) & ($1,1,0$) & ($0,0,1$) & ($0,0,0,0,0,0,0,1,0$) 
 \\ \hline

($0,0,0$) & ($0,0,0$) & ($1,1,0$) & ($0,0,0$) & ($0,0,1$) & ($0,0,0,0,0,0,0,0,1$) 
 \\ \hline

($0,0,0$) & ($1,1,0$) & ($0,0,0$) & ($0,0,0$) & ($0,0,1$) & ($0,0,0,0,0,0,0,0,0$) 
 \\ \hline

($1,1,0$) & ($0,0,0$) & ($0,0,0$) & ($0,0,0$) & ($0,0,0$) & ($0,0,0,0,0,0,1,0,0$) 
 \\ \hline \hline

($0,0,0$) & ($0,0,0$) & ($0,0,1$) & ($0,0,1$) & ($0,0,1$) & ($0,0,0,0,0,0,1,0,0$) 
 \\ \hline 

($0,0,0$) & ($0,0,1$) & ($0,0,0$) & ($0,0,1$) & ($0,0,1$) & ($0,0,0,0,0,0,1,0,1$) 
 \\ \hline 

($0,0,0$) & ($0,0,1$) & ($0,0,1$) & ($0,0,0$) & ($0,0,1$) & ($0,0,0,0,0,0,1,1,0$) 
 \\ \hline 

($0,0,1$) & ($0,0,0$) & ($0,0,0$) & ($0,0,1$) & ($0,0,2$) & ($0,0,0,0,0,0,0,0,1$) 
 \\ \hline 

($0,0,1$) & ($0,0,0$) & ($0,0,1$) & ($0,0,0$) & ($0,0,2$) & ($0,0,0,0,0,0,0,1,0$) 
 \\ \hline 

($0,0,1$) & ($0,0,1$) & ($0,0,0$) & ($0,0,0$) & ($0,0,2$) & ($0,0,0,0,0,0,0,1,1$) 
 \\ 
\end{tabular}
\end{center}
\caption{A set of the solutions for a semi-realistic model
$U(3)\times U(2)^2 \times U(1)$ arising from the D3 brane sector.    } 
\end{table}

However, before doing so, we would like  to remark that tadpole
cancellation  constraints do not allow for  a realistic gauge
structure arising from
 D3-branes at the
$\IC^3/\IZ_3$ orbifold points. Solutions with the maximal allowed
number of D3-branes placed at  $\IZ_3$ fixed points correspond to six
D3-branes at such orbifold points.
However, as a result of strong constraints from tadpole cancellation
conditions, the  gauge structure arising form these D3-branes is $U(2)^3$.
Therefore,  N=1 supersymmetric solutions of the 
$\IZ_6$ orientifold model do not reproduce the 
local configurations  with realistic gauge group
structure at $\IZ_3$ fixed points, such as studied in
\cite{Aldazabal}.  However, in Subsection \ref{anti} we show how the
introduction of
anti-branes, which break supersymmetry, allows for a realistic
gauge groups  arising from  D3-branes at $\IZ_3$
orbifold points.

\subsection{A semi-realistic model}
\label{semi}
In this section we shall focus on the model with the semi-realistic gauge
group  structure. This model can be obtained with the following choice of
values for the integer parameters defined in Subsection \ref{smodels},
\beqa
N_1=2&,&\ \ N_2=3, \ \ N_3=4, \ \ N_4=1, \ \ N_5=0, \ \ N_6=2;\nonumber \\
M_1=1&,&\ \ M_2=0, \ \ M_3=0, \ \ M_4=2, \ \ M_5=3, \ \ M_6=2;\nonumber \\
j_1=1&,&\ \ j_2=1, \ \ j_3=0; \nonumber \\
m_1=0&,&\ \ m_2=0, \ \ m_3=1,
\eeqa
while $i_n=k_n=l_n=x_n=0$. Namely, the model has 8 D3-branes at the origin, 
8 D7$_3$-branes at $p=0$, 12 D7$_3$ branes at $p=1$ and 12 at its orientifold 
image $p=-1$. In addition, there are 2 $D3$-branes at the point $(0,1,1)$ and  
its three images. There is also one order-three Wilson line acting on the
first complex plane. 

The gauge group on the D7$_3$-brane sector at $p=1$, before the Wilson
line projection is imposed, is [with grouping  of the integer entries as
$(N_1,N_2,N_3,N_4,N_5,N_6)$]:
\beq
G_{\theta,7_3,1}=U(2)\times  U(3)\times U(4) \times U(1)\times U(2)\; .
\eeq
The Wilson line breaks $U(4)$ to $U(3)\times U(1)'$ and the first $U(2)$
to $U(1)\times U(1)''$, and further breaks $U(1)'\times U(1)''$ to the
diagonal combination $U(1)_{diag}$. The remaining group is
\beq
G_{\theta,7_3,1}^{\rm Wilson} = U(2)\times U(3)\times U(3)\times
U(1) \times U(1)\times U(1)_{diag}.
\eeq

The $7_37_3$ spectrum before the Wilson line projection is 
\beqa
2\; \times\; [\;
({\bf 2},{\bf {\ov 3}},{\bf 1},{\bf 1})_{[1,-1,0,0,0]} + 
({\bf 1},{\bf 3},{\bf {\ov 4}},{\bf 1})_{[0,1,-1,0,0]} +
({\bf 1},{\bf 1},{\bf 4},{\bf 1})_{[0,0,1,-1,0]} + 
({\bf {\ov 2}},{\bf 1},{\bf 1},{\bf 2})_{[-1,0,0,0,1]} \; ] \; +
\nonumber \\
+ \; ({\bf {\ov 2}},{\bf 1},{\bf 4},{\bf 1})_{[-1,0,1,0,0]} + 
({\bf 1},{\bf {\ov 3}},{\bf 1},{\bf 1})_{[0,-1,0,1,0]} +
({\bf 1},{\bf 1},{\bf 1},{\bf 2})_{[0,0,0,-1,1]} + 
({\bf 1},{\bf 3},{\bf 1},{\bf {\ov 2}})_{[0,1,0,0,-1]} \; , 
\eeqa
where we put a bar over the fundamentals of $SU(2)$ which have $-1$ charge under
the corresponding $U(1)$ in $U(2)$.

After the Wilson line projection, the spectrum in this sector is
\begin{equation}
\begin{array}{cccc}
A_1^{(7)}, B_1^{(7)}: & 
2&\times & ({\bf 2},{\ov {\bf 3}},{\bf 1})_{[1,-1,0,0,0,0]}; \\
A_2^{(7)}, B_2^{(7)}: 
& 2&\times & ({\bf 1},{\bf 3},{\ov{\bf 3}})_{[0,1,-1,0,0,0]};\\
A_3^{(7)}, B_3^{(7)}:
& 2&\times & ({\bf 1}, {\bf 1}, {\bf 3})_{[0,0,1,-1,0,0]};\\
A_6^{(7)}, B_6^{(7)}:
& 2&\times & ({\ov{\bf 2}},{\bf 1},{\bf 1})_{[-1,0,0,0,1,0]};\\
C_2^{(7)} :
& & & ({\bf 1}, {\bf 3}, {\bf 1})_{[0,1,0,0,-1,0]}; \\
C_3^{(7)}:
& & & ({\ov{\bf 2}},{\bf 1},{\bf 3})_{[-1,0,1,0,0,0]};\\
C_4^{(7)}:
& & & ({\bf 1}, {\ov{\bf 3}}, {\bf 1})_{[0,-1,0,1,0,0]};\\
C_6^{(7)}
& & & ({\bf 1}, {\bf 1}, {\bf 1})_{[0,0,0,-1,1,0]}\; ,\\
\end{array}
\end{equation}
where the representations are under $SU(2)\times SU(3)\times SU(3)$, and
the subscripts correspond to $U(1)$ charges. The superscripts $(7)$ 
 denote the states arising in a D7-brane sector.

The gauge group on the $D3$-sector at the point $(0,0,1)$ (and its
orientifold image) is of the form:
\beq
G_{\theta,3,(0,0,1)}=U(1)\times  U(2)\times U(3)\times U(2)\; .
\eeq
The spectrum of chiral multiplets in this $33$ sector is
\begin{equation}
\begin{array}{cccc}
A_4^{(3)}, B_4^{(3)}:
&2&\times & ({\bf 2},{\ov{\bf 3}}, {\bf 1})_{[0,1,-1,0]};\nonumber\\
A_5^{(3)}, B_5^{(3)}:
&2&\times & ({\bf 1},{\bf 3},{\ov{\bf 2}})_{[0,0,1,-1]};\nonumber\\
A_6^{(3)}, B_6^{(3)}:
&2&\times & ({\bf 1}, {\bf 1}, {\bf 2})_{[-1,0,0,1]};\nonumber\\
C_1^{(3)}
& & & ({\bf 1},{\ov{\bf 3}},{\bf 1})_{[1,0,-1,0]}; \nonumber \\
C_6^{(3)}
& & & ({\ov{\bf 2}},{\bf 1},{\bf 2})_{[0,-1,0,1]}\; .\nonumber
\end{array}
\end{equation}
The superscripts $(3)$ denote the corresponding fields arise from the
D3-brane sector.

In the $37_3$ and $7_33$ sectors, we obtain
\beqa
{\bf 37_3} \;\;
& D_1^{(37)}: \;\; &({\bf 1},{\bf 1},{\bf 1})_{[1,0,0,0]} \times 
({\bf 1},{\bf {\ov 3}},{\bf 1})_{[0,-1,0,0,0,0]} \; ; \nonumber\\
& D_5^{(37)}: \;\; &
({\bf 1},{\bf 3},{\bf 1})_{[0,0,1,0]} \times 
({\bf 1},{\bf 1},{\bf 1})_{[0,0,0,0,-1,0]} \; ; \nonumber \\
&{\tilde D}_5^{(37)}: \;\; & 
({\bf 1},{\bf 3},{\bf 1})_{[0,0,1,0]} \times 
({\bf 1},{\bf 1},{\bf 1})_{[0,0,0,0,0,-1]} \; ; \nonumber \\
& D_6^{(37)}:\;\; &
({\bf 1},{\bf 1},{\bf 2})_{[0,0,0,1]} \times 
({\bf {\ov 2}},{\bf 1},{\bf 1})_{[-1,0,0,0,0,0]} \; ; \nonumber \\
{\bf 7_33} \;\;
& E_3^{(73)}:\;\; &
({\bf {\ov 2}},{\bf 1},{\bf 1})_{[0,-1,0,0]} \times
({\bf 1},{\bf 1},{\bf 3})_{[0,0,1,0,0,0]} \; ; \nonumber \\
&{\tilde E}_3^{(73)}:\;\; &
({\bf {\ov 2}},{\bf 1},{\bf 1})_{[0,-1,0,0]} \times
({\bf 1},{\bf 1},{\bf 1})_{[0,0,0,0,0,1]} \; ; \nonumber \\
& E_4^{(73)}:\;\; &
({\bf 1},{\bf {\ov 3}},{\bf 1})_{[0,0,-1,0]} \times
({\bf 1},{\bf 1},{\bf 1})_{[0,0,0,1,0,0]} \; ; \nonumber \\
& E_6^{(73)}:\;\; &
({\bf 1},{\bf 1},{\bf 1})_{[-1,0,0,0]} \times 
({\bf 1},{\bf 1},{\bf 1})_{[0,0,0,0,1,0]} \; ; \nonumber \\
& {\tilde E}_6^{(73)}:\;\; &
({\bf 1},{\bf 1},{\bf 1})_{[-1,0,0,0]} \times 
({\bf 1},{\bf 1},{\bf 1})_{[0,0,0,0,0,1]} \; , \nonumber 
\eeqa
where the two parts for each state give the quantum numbers under the D3-
and D7$_3$-brane gauge groups.

The D3-branes at the point $(0,1,1)$ (and its images) produce a gauge
group $G_{\theta,3,(0,1,1)}=U(1)\times U(1)$. In the 33 sector, we obtain the multiplets
\beqa
A_2^{(3')}, B_2^{(3')}, C_2^{(3')}:\;\; & 3\; \times\; {\bf 1}_{[-1,1]}\; ,
\eeqa
where we use a prime to distinguish this D3-brane sector from the previous
one. In the $37_3$ and $7_33$ sectors we obtain chiral multiplets
\beqa
{\bf 37_3} \;\;& 
D_1^{(3'7)}:\;\; &
{\bf 1}_{[1,0]}\times ({\bf 1},{\bf 1},{\bf {\ov 3}})_{[0,0,-1,0,0,0]} 
\; ; \nonumber\\
&{\tilde D}_1^{(3'7)}:\;\; &
{\bf 1}_{[1,0]}\times ({\bf 1},{\bf 1},{\bf 1})_{[0,0,0,0,0,0,-1]} 
\; ; \nonumber \\
&G_1^{(3'7)}:\;\; &
{\bf 1}_{[1,0]}\times ({\bf 1},{\bf 1},{\bf 1})_{[0,0,0,0,0,-1,0]} 
\; ; \nonumber \\
&{\tilde G}_1^{(3'7)}:\;\; &
{\bf 1}_{[1,0]}\times ({\bf 1},{\bf 1},{\bf 1})_{[0,0,0,0,0,0,-1]} 
\; ; \nonumber \\
& D_3^{(3'7)}:\;\; &
{\bf 1}_{[0,1]}\times ({\bf 1},{\bf{\ov 3}},{\bf 1})_{[0,-1,0,0,0,0]}
\; ; \nonumber \\
{\bf 7_33} \;\;& 
E_1^{(73')}:\;\; &
{\bf 1}_{[-1,0]}\times ({\bf 2},{\bf 1},{\bf 1})_{[1,0,0,0,0,0]} 
\; ; \nonumber \\
& E_3^{(73')}:\;\; &
{\bf 1}_{[0,-1]}\times ({\bf 1},{\bf 1},{\bf 3})_{[0,0,1,0,0,0]} 
\; ; \nonumber \\
& {\tilde E}_3^{(73')}:\;\; &
{\bf 1}_{[0,-1]}\times ({\bf 1},{\bf 1},{\bf 1})_{[0,0,0,0,0,1]}
\; ; \nonumber \\
& E_4^{(73')}:\;\; &
{\bf 1}_{[-1,0]}\times ({\bf 1},{\bf 1},{\bf 1})_{[0,0,0,1,0,0]} 
\; ; \nonumber \\
& E_6^{(73')}:\;\; &
{\bf 1}_{[0,-1]}\times ({\bf 1},{\bf 1},{\bf 1})_{[0,0,0,0,1,0]} 
\; ; \nonumber \\
& {\tilde E}_6^{(73')}:\;\; &
{\bf 1}_{[0,-1]}\times ({\bf 1},{\bf 1},{\bf 1})_{[0,0,0,0,0,1]}\; .
\eeqa

To complete the model, we now turn to  the D$7_3$-branes placed at $p=0$ 
and the D3-branes placed at the origin. Each of them gives rise to the
gauge group 
\beqa
G_{\theta, 7_3, 0}=G_{\theta, 3, (0,0,0)}=U(2)\times U(2) \; .
\eeqa
In the $7_37_3$ sector we obtain chiral multiplets in the representation
\beqa
2\;\times\; [\; ({\bf 1},{\bf 1})_{[0,2]} + ({\bf 1},{\bf 2})_{[-2,0]} \;
] + ({\bf {\ov 2}},{\bf 2})_{[-1,1]} \; .
\eeqa
In the $33$ sector we obtain the same type of spectrum. The $37_3$ and
$7_33$ are related by the orientifold projection, so it suffices to
compute just the former, which produces fields
\beqa
({\bf {\ov 2}},{\bf 1})_{[-1,0]}\times ({\bf {\ov 2}},{\bf 1})_{[-1,0]} + 
({\bf 1},{\bf 2})_{[0,1]}\times ({\bf 1},{\bf 2})_{[0,1]} \; .
\eeqa

It is easy to check that all cubic non-Abelian anomalies cancel.

The full superpotential for fields arising from branes at $z_3=1$ is given
by
\beqa
W  =
& A_6^{(7)} B_1^{(7)} C_2^{(7)} - B_6^{(7)} A_1^{(7)} C_2^{(7)} +
A_1^{(7)} B_2^{(7)} C_3^{(7)} - B_1^{(7)} A_2^{(7)} C_3^{(7)} + 
A_2^{(7)} B_3^{(7)} C_4^{(7)} - B_2^{(7)} A_3^{(7)} C_4^{(7)}+ 
& \nonumber\\
 & +A_5^{(3)} B_6^{(3)} C_1^{(3)} - B_5^{(3)} A_6^{(3)} C_1^{(3)} +
A_4^{(3)} B_5^{(3)} C_6^{(3)} - B_4^{(3)} A_5^{(3)} C_6^{(3)} + & 
\nonumber\\
&+ D_5^{(37)} E_6^{(73)} C_1^{(3)} + E_4^{(73)} D_5^{(37)} C_6^{(7)} +
E_6^{(73)} D_1^{(37)} C_2^{(7)}+ & \nonumber \\
&+ C_2^{(3')} D_1^{(3'7)} E_3^{(73')} + C_2^{(3')} {\tilde D}_1^{(3'7)}
{\tilde E}_3^{(73')} +  C_2^{(3')} G_1^{(3'7)} E_6^{(73')} +  C_2^{(3')}
{\tilde G}_1^{(3'7)} {\tilde E}_6^{(73')} \; . &
\eeqa

As usual in type IIB orientifold models, mixed $U(1)$ triangle anomalies
do not vanish, but are canceled by a generalized Green-Schwarz mechanism
mediated by closed string twisted modes \cite{iru} (see \cite{sagnan} for 
the six-dimensional version of this effect). The anomalous $U(1)$'s gain a
mass of the order of the string scale \cite{poppitz}, and disappear from
the low-energy physics. It is therefore important to determine the
non-anomalous (and therefore light) $U(1)$'s in the model.

The non-anomalous U(1)'s can be found by directly computing the
matrix of
mixed $U(1)$-non-Abelian anomalies. We order the gauge factors as they
arise in
\beq
G_{\q, 7_3, 1}^{Wilson} \times G_{\q,3,(0,0,1)} \times
G_{\q,3,(0,1,1)}\; ,  
\eeq
and denote each $U(1)$ generator by $Q_i$, i=1,...,12. Denoting by
$A_{ij}$ the mixed $Q_i-SU(N_j)^2$ anomaly, non-anomalous $U(1)$'s
correspond to zero  eigenvalue modes in this matrix. Namely, linear
combinations 
$Q^{(k)} = \sum_i \frac{c^{(k)}_i}{N_i} Q_i$ satisfying $\sum_i
\frac{c^{(k)}_i}{N_i} A_{ij}=0$. In our model there are four independent
non-anomalous $U(1)$'s, whose 12-dimensional vector of coefficients
$(c_i)$ are given by
\beq
\begin{array}{c}
c^{(1)}=(-1,0,0,1,0,0;1,-1,0,0;0,0)\; ; \\
c^{(2)}=(0,0,0,0,1,1;0,0,0,-1;1,1)\; ; \\
c^{(3)}=(1,1,0,0,1,0;-1,1,0,0;0,0)\; ; \\
c^{(4)}=(1,1,1,0,0,0;0,1,1,1;0,0)\; . 
\end{array}
\eeq
Clearly, any linear combination of these is also non-anomalous. For instance, 
there exists one non-anomalous $U(1)$ arising only from D7$_3$-branes at
$z_3=+1$, without any D3-brane component. Its vector of components is
obtained by adding the first and third vectors above, giving
\beq
\begin{array}{c}
(0,1,0,1,1,0;0,0,0,0;0,0)\; , 
\end{array}
\eeq
which corresponds to $Q = \textstyle{2\over 3} Q_2+ Q_4 + Q_5$. 

In an attempt to identify possible candidates for $U(1)_Y$ hypercharge in 
the model, we consider the following scenario. Fields $A_4,B_4$ are taken
to be quark singlets, while $A_5, B_5$ are identified as quark doublets,
$A_6, B_6$ are lepton doublets and $C_6$ is the Higgs. Therefore
the
sector of D3-branes at the point $(0,0,1)$ gives rise to a left-right
symmetric semi-realistic sector with group $SU(3)_{color}\times SU(2)_L\times
SU(2)_R$, and two quark-lepton families. In order to obtain correct
hypercharge assignments after
breaking of $SU(2)_R$, we need a correct ($B$-$L$)- charge $Q_{B-L}$ to form
the linear combination $Q_{Y}=Q_{B-L}+2I_R$. Defining  
\beq
Q_{B-L}=a\times c^{(1)} + b\times c^{(2)}+ c\times c^{(3)} + d\times c^{(4)}\,  . 
\eeq
and requiring that $Q_{B-L}$ charge assignments are correct, one arrives
at 
the following two constraints
\beq
a=b+c \ \ ; \ \  d=3b-2. 
\eeq
A simple solution is $b=c=0$, leading to $Q_{B-L}=-2\times c^{(4)}$, which
can be cast in the form
\beqa
Q_{B-L} = (\frac 12 Q_1 + \frac 13 Q_2 + \frac 13 Q_3) + (\frac 12 Q_8 + 
\frac 13 Q_9 + \frac 12 Q_{10}) \; , 
\eeqa
where the first bracket arises from D7$_3$-brane gauge factors, and the
second one from D3-branes gauge factors located at $(0,0,1)$. Notice that,
as opposed to the realistic models in \cite{Aldazabal,aiq}, the 
hypercharge in this model arises
from $U(1)$ linear combinations of different brane sectors. It would be
interesting to gain a better understanding of the origin  of such
non-anomalous  $U(1)$'s as ($B$-$L$) and/or  hypercharge candidates 
and their phenomenological implications.

We refrain from entering a more detailed discussion of the
phenomenological properties of this mode, e.g., the 
charge assignments of fields in other sectors, or the study of
superpotential couplings,   which is a subject of further studies
\cite{CLUW}. In the next Subsection we turn to the discussion of 
orientifold models with broken supersymmetry.

\subsection{A semi-realistic model with anti-branes}
\label{anti}
It is possible to see that the main obstruction to obtain three-family
models with realistic gauge groups in the framework we have described is
the limited amount of available branes. For instance, following the
approach in \cite{Aldazabal}, one may try to obtain such a realistic
sector from a set of six D3-branes at one of the $\IC^3/\IZ_3$ orbifold
points. Since such points have images under $\theta^3$ and $\Omega_3$,
 the number of
D3-branes involved is 24. Counting also the D3-branes at $(0,0,0)$ whose
minimum number is 8, we saturate the total number of available 32
D3-branes. However, all models built in this manner with the requirement that the set of six D3-branes at one of the $\IC^3/\IZ_3$ orbifold gives rise to realistic gauge structure, i.e., $SU(3)\times SU(2)\times U(1)$,  contain non-zero
tadpoles at some of the additional $\IC^3/\IZ_3$ orbifold points,
arising from D7$_3$-brane disks. Since all 32 D3-branes have been already
placed at other points, the tadpoles remain uncancelled, rendering such
models inconsistent.

The introduction of anti-branes in the construction of orientifolds
\cite{antibranes} (see also \cite{aiq,othernonsusy} for further
developments) allows one  to build models where, e.g., the number of
D3-branes
$N$ is larger than 32, the excess of untwisted charge being compensated by
$N-32$ anti-D3-branes (denoted \Dthree-branes). In this subsection we
would like to briefly discuss the construction of a semi-realistic
orientifold, with Pati-Salam gauge group $SU(3)_{color}\times SU(2)_L\times
SU(2)_R\times U(1)_{B-L}$ and three families of quarks and leptons with
correct gauge quantum numbers. This sector arises from a set of D3-branes
sitting at a $\IC^3/\IZ_3$ orbifold point, and is identical to those
studied in \cite{Aldazabal}.

We start by placing 20 D7$_3$-branes at $z_3=0$ with
\beqa
\gamma_{\theta,7_3,0}=\diag(e^{\pi i \frac 16} \id_4, e^{\pi i \frac 36}
\id_2, e^{\pi i \frac 56} \id_4, e^{\pi i \frac 76} \id_4, e^{\pi i \frac
96} \id_2, e^{\pi i\frac{11}{6}} \id_4) \; . 
\eeqa
This choice differs from (\ref{chorigin}) only by a traceless piece, so
the analysis is similar to that described in Subsection \ref{smodels}. In
particular, all crosscap tadpoles are canceled at points $(m,n,0)$,
except at the origin. In order to cancel the latter, we choose to 
introduce four \Dthree-branes with
\beqa
\gamma_{\theta,{\ov 3},(0,0,0)}=\diag(e^{\pi i \frac 36} \id_2, 
e^{\pi i \frac 96} \id_2) \; . 
\label{choanti}
\eeqa
\Dthree-branes carry RR charges opposite to those of D3-branes, hence
their contribution to tadpoles is opposite to that of a set of D3-branes
with the same Chan-Paton matrix. Using this fact, the above choice is seen
to cancel the tadpole at $(0,0,0)$. 

Up to this point we have introduced 20 D7$_3$-branes and 4 \Dthree-branes,
so cancellation of untwisted tadpoles requires the introduction of 12
additional D7$_3$-branes, and 36 D3-branes, split symmetrically between
the points at $z_3=1$ and its $\Omega_3$ images. The discussion is
analogous to that in Subsection \ref{smodels}, differing only in the total
number of branes allowed. 

We locate six D7$_3$-branes at $z_3=1$ (and the remaining 6 at its
$\Omega_3$-image), with  
\beqa
\gamma_{\theta,7_3,1} & = &\diag(e^{\pi i \frac 16}, e^{\pi i \frac 36},
e^{\pi i \frac 56}, e^{\pi i \frac 76}, e^{\pi i \frac 96}, e^{\pi
i\frac{11}{6}}) \; , 
\eeqa
that is $N_1=N_2=N_3=N_4=N_5=N_6=2$ in (\ref{enes}). Since this matrix is
traceless, we may choose to place no D3-branes at $(0,0,1)$, i.e.
$M_i=0$ in (\ref{emes}). In order to have nontrivial sectors at the
$\IC^3/\IZ_3$ orbifold points, we introduce an order three Wilson line
along the first plane on the D7$_3$-branes, corresponding to $m_1=m_3=0$,
$m_2=1$ in (\ref{w1}), so in the basis diagonalizing 
$\gamma_{\theta^2,7_3,+1}$ and $\gamma_{W_1,7_3,+1}$ it reads
\beqa
{\tilde \gamma}_{W_1,7_3,+1} & = & \diag(1,e^{2\pi i\frac 13}, 1,1,
e^{2\pi i\frac 23},1) \; . 
\eeqa
Finally, we turn to the choice of location and Chan-Paton factors
for D3-branes (\ref{multiplechoice}), with the modified constraint
$\sum_n (2i_n + 2j_n + 2k_n + 2 l_n)=18$.  Fixed points of the form
$(0,n,1)$ do not feel the Wilson line, and their tadpoles cancel
without the introduction of D3-branes sitting at them, hence $j_n=0$. 
At the remaining points, the D3-brane Chan-Paton trace should be $-1$, so
we may choose $k_1=l_1=1$, $k_2=k_3=l_2=l_3=0$, $i_1=3$, $i_2=i_3=2$.

The computation of the open string massless spectrum is lengthy, but
straightforward. In any event, the important aspect we would like to point
out is that the local behavior near the fixed point $(1,0,1)$ (and its
images) is exactly as in the models in \cite{aiq,Aldazabal}, hence this
D3-brane sector leads automatically to a three-family $SU(3)_{color}\times
SU(2)_L\times SU(2)_R\times U(1)_{B-L}$. The  $U(1)_{B-L}$ arises
from the unique non-anomalous linear combination of $U(1)$'s, purely in
the D3-brane sector, and whose charge is given by
\beqa
Q_{B-L} = -2(\frac 13 Q_{U(3)}+\frac 12 Q_{U(2)} + \frac 12 Q_{U(2)}).
\eeqa
Also note that, even though the
full model is non-supersymmetric due to the presence of the anti-branes,
the latter sit at $z_3=0$, i.e. are separated from the semi-realistic
sector, which enjoys $N=1$ supersymmetry, only broken at higher orders
by the hidden anti-brane sector.

The spectrum charged under gauge factors in this sector of D3-branes is
\beqa
& SU(3)\times SU(2)_L\times SU(2)_R \times U(1)_{B-L}  & \nonumber \\
{\bf 33}\; : \quad & 3\; \times\; [\;
(3,2,1)_{1/3} + (1,2,2)_{0} + ({\ov 3},1,1)_{-1/3} \;] \; ; \nonumber \\
{\bf 37_3}\; :\quad & 3\;\times\; [\; (3,1,1)_{-2/3} + (1,2,1)_{-1} \;]
\; ; \nonumber\\
{\bf 7_33}\; :\quad & 3\;\times\; [\; (1,1,2)_{1} + ({\ov 3},1,1)_{2/3} \;]\; .
\eeqa
The three copies of fields in the mixed sector actually differ in their
charges under the D7$_3$-brane gauge group $U(1)^3$, not shown here.
In fact, we expect these additional $U(1)$'s to be anomalous, and
therefore broken and not present at low energies. We will not pursue their
discussion here.

Finally note that the model we have just constructed is in principle
unstable against moving some of the D3-branes at $(1,0,1)$ and
annihilating them with the \Dthree-branes at the origin. This instability
is identical to that in the models in \cite{aiq}, but it can be avoided
in more involved constructions \cite{Aldazabal}. In any event, our
purpose was to illustrate how the construction of realistic models using
the $\IZ_6$ orientifold is possible and relatively easy once the
restriction of supersymmetry is relaxed.  

\section{Conclusions}
\label{conclusions}

In this paper we set out to develop techniques to construct consistent N=1 
supersymmetric four-dimensional solutions of open (Type I) string theory
compactified on $\IZ_N$ and $\IZ_M\times \IZ_N$ symmetric orientifolds,
{\it with general (continuous and discrete) Wilson lines}. In particular,
our approach advances the techniques beyond the special examples of discrete 
\cite{afiv,wilsonlinemodel,CPW,lykken} and continuous \cite{afiv,ibanez,cl} 
Wilson lines. 
We have provided explicit solutions for the algebraic
consistency conditions for discrete and continuous Wilson lines along
complex planes twisted by an order six action. We have also studied the
tadpole consistency conditions on the $\IZ_6$ orientifold with general
Wilson lines, and found explicit solution to this set of constraints,
leading to interesting N=1 four-dimensional string vacua. And we would like to
emphasize that our techniques in the construction of Wilson
lines are general and applicable to other models.

In constructing models with Wilson lines 
we have heavily employed the geometrical interpretation of discrete Wilson 
line solutions in the T-dual picture as sets of branes located at the orbifold 
fixed points. In the T-dual  
picture the algebraic consistency conditions become constraints   
for the locations of these branes that are consistent with the orbifold 
and orientifold symmetries.
We have also shown that the tadpole cancellation conditions for such T-dual
models are related by a discrete Fourier transform. This useful trick
allows one to obtain tadpoles in T-dual models without the need of directly
computing them, and illuminates the detailed relation between brane
positions and Wilson lines in orbifold models.

The second major motivation for advancing this program is phenomenological.
Models with discrete Wilson lines naturally provide smaller gauge group 
structures (associated with the set of branes located at the orbifold
points), and thus could potentially lead to gauge sectors with the
ingredients of the standard model. However, previous four-dimensional N=1 
supersymmetric Type I models (with relatively simple configurations of
brane locations and Wilson lines) yield gauge groups that cannot directly 
provide a $SU(3)_{color}$ candidate (see, e.g., \cite{CPW,unpub} and
references therein). In addition, most of the $U(1)$ factors were anomalous
and therefore massive, leading to a generic difficulty in obtaining
$U(1)_{Y}$-hypercharge candidates for the standard model \cite{unpub}.  
The particular $\IZ_6$ orientifold, explored in this paper, could potentially
provide the candidate group structure for both $SU(3)_{color}$ as well as 
non-anomalous $U(1)_Y$. 
The latter motivation was originally based on the observation
\cite{Aldazabal} that branes at {\em orbifold} points, which are 
not fixed under the orientifold projection, may allow for appearance of
non-anomalous $U(1)$'s, that could play the role of hypercharge. 

The systematic exploration of N=1 supersymmetric discrete Wilson line 
solutions yields a surprisingly small number of non-equivalent classes of
models. This result is due to the extremely tight tadpole consistency
conditions for N=1 supersymmetric models, as well as the large symmetry
group of the underlying geometry $T^6/\IZ_6$, which we exploit extensively 
to identify different equivalence classes. In particular, N=1 supersymmetry 
constraints do not allow us to locate six (or more) branes at the $\IZ_3$ 
orbifold points and obtain realistic gauge structure from these 
D3 branes in the model, even though 
this is a configuration that has the potential for yielding
three-family sectors with the standard model (or left-right
symmetric) gauge group. However, as we demonstrated in subsection
\ref{anti},
such configurations are possible if one allows for the introduction of
anti-branes in the model, which lead to the breakdown of supersymmetry in
the anti-brane sector.

The  upshot of the exploration of N=1 supersymmetric models, which contain 
sectors with $SU(3)$ as a $SU(3)_{color}$ candidate, is the existence of
 only two  inequivalent classes of such  solutions. We provide the complete spectrum and
trilinear superpotential couplings for one class (the second equivalence  
class would yield an equivalent ``observable sector'' gauge group and 
spectrum). The model allows for the gauge group assignment: $SU(3)_{color}
\times SU(2)_L\times SU(2)_R\times U(1)_{B-L}$, with the non-Abelian part
arising from branes located at the $\IZ_6$ orbifold fixed point. 
Interestingly, the non-anomalous Abelian factor $U(1)_{B-L}$ is however a
combination of $U(1)$ factors arising from different sets of branes, namely 
D3- and D7-branes. Unfortunately, the above particular gauge group 
assignment has a major flaw, since it leads to only {  two}  sets of
quark families. Nevertheless, other observable sector gauge group
assignments are possible, and a more general phenomenological exploration
of this model is under way \cite{CLUW}.

\acknowledgments

We would like to thank P. Langacker for many discussions, suggestions
regarding the manuscript and for collaboration on related topics.
We also benefited from discussions with G. Aldazabal, L.~E.~Ib\'a\~nez, M.
Pl\" umacher, F.~Quevedo and R. Rabad\'an. 
We would like to thank SISSA, Trieste, Italy (M.C.) CAMTP, Maribor, 
Slovenia (M.C.), the Department of Physics and Astronomy, Rutgers
University (M.C.), the Department of Physics and Astronomy of the University
of Pennsylvania (A.U., J.W.), and the Centro At\'omico Bariloche, 
Argentina (A.U.) for support and hospitality during the completion of the
work. The work was supported in part by U.S.\ Department of Energy Grant
No.~DOE-EY-76-02-3071 (M.C.), in part by the University of Pennsylvania
Research Foundation award (M.C.) and the NATO Linkage grant 97061 (M.C.).
J.W. is supported by Department of Energy Grant No.~DE-AC02-76CH03000.

\clearpage

\appendix{\bf Appendix:  
T-duality and Fourier transformation of tadpole conditions}

In this Appendix we show that tadpole cancellation conditions of
orientifold models related by T-duality are related by a discrete
Fourier-transform. The basic motivation for the proposal is that T-duality
maps twisted sectors of one model to linear combinations of twisted
sectors in the T-dual, with coefficients defining precisely a discrete
Fourier transform. Hence orientifold twisted charges in a model and its
T-dual must be related by such a discrete Fourier transform. Finally,
consistent coupling of D-branes to twisted fields in a model and its
T-dual implies that D-brane twisted charges, e.g., their contribution to
twisted tadpoles, must also be related in this way.

In order to be concrete, we focus  on the specific example of the
$T^6/\IZ_3$ orientifold, even though the technique is valid in general.
Let us consider a $T^6/\IZ_3$ type IIB orbifold modded out by $\Omega$
\cite{ABPSS}, and let us introduce general order three Wilson lines along
the three complex planes and acting on the D9-branes through matrices
$\gamma_{W_i,9}$. The twisted tadpole conditions at a $\theta$-fixed 
point, generally denoted $(m_1,m_2,m_3)$, with $m_i=0, \pm 1$, read
\beqa
\begin{array}{lcl}
(m_1,m_2,m_3) & \quad &\Tr (\gamma_{\theta,9} \gamma_{W_1,9}^{m_1} 
\gamma_{W_2,9}^{m_2} \gamma_{W_3,9}^{m_3})=-4\; ; \\
& & \Tr (\gamma_{\theta^2,9} \gamma_{W_1,9}^{2m_1} \gamma_{W_2,9}^{2m_2}
\gamma_{W_3,9}^{2m_3})=-4\; . \\
\end{array}
\label{tadone}
\eeqa
It is convenient to make them a bit more explicit. Since all matrices commute 
and are of order three, they can be diagonalized, with eigenvalues $1,
e^{2\pi i/3}, e^{2\pi i\frac 23}$. Let us denote by $n_{r,s_1,s_2,s_3}$
the number of entries with eigenvalues $e^{2\pi i r/3}$, $e^{2\pi i
s_a/3}$, in $\gamma_{\theta,9}$, $\gamma_{W_a,9}$, respectively. The
tadpole conditions read
\beqa
\begin{array}{lcl}
(m_1,m_2,m_3) & \quad & \sum_{r,s_1,s_2,s_3} e^{2\pi i r/3} 
e^{2\pi i\,m_1 s_1/3} e^{2\pi i\, m_2 s_2/3}  e^{2\pi i \, m_3 s_3/3}
n_{r,s_1,s_2,s_3} = -4 \; ;\\
       & & \sum_{r,s_1,s_2,s_3} e^{2\pi i 2r/3} e^{2\pi i\, 2m_1 s_1/3}
e^{2\pi i\, 2m_2s_2/3} e^{2\pi i \, 2m_3s_3/3} n_{r,s_1,s_2,s_3} = -4\; . \\
\end{array}
\label{tadtwo}
\eeqa
Performing a T-duality along all compact directions, we get a $T^6/\IZ_3$
type IIB orbifold modded out by $\Omega (-1)^{F_L} R$, where $R$ reflects
all internal dimensions (see \cite{lykken} for a detailed description of 
this model). This T-dual model contains D3$'$-branes (we use primes to
denote D-branes in the T-dual picture) which are located at the 
27 $\theta$-fixed points, denoted $(t_1,t_2,t_3)$, $t_a=0,\pm 1$. By the
usual T-duality correspondence between Wilson line eigenvalues and D-brane
positions, D9-branes with Wilson line eigenvalues defined by $s_i$ map
to D3$'$-branes sitting at $(t_1,t_2,t_3)=(s_1,s_2,s_3)$, hence
$\gamma_{\theta,3,(t_1,t_2,t_3)}$ has $n_{r,t_1,t_2,t_3}$ eigenvalues
$e^{2\pi ir/3}$. This completely specifies the T-dual model in terms of
the original one. 

Our purpose now is to perform the discrete Fourier transform in the
tadpole cancellation conditions (\ref{tadtwo}), and show that they
correspond to the tadpole cancellation conditions in the T-dual
description. For instance, for the $\theta$-twisted tadpole equation, 
multiplying the first equation in (\ref{tadtwo}) by $e^{-2\pi i \,m_1t_1/3} 
e^{-2\pi i\, m_2t_2/3}  e^{-2\pi i\, m_3t_3/3}$ and summing over
$m_1,m_2,m_3$, we get
\beqa
27 \sum_{r,s_1,s_2,s_3} e^{2\pi i r/3} \delta_{s_1,t_1} \delta_{s_2,t_2}
\delta_{s_3,t_3} n_{r,s_1,s_2,s_3}= -4 \times 27 \times \delta_{t_1,0}
\delta_{t_2,0} \delta_{t_3,0}\; .
\eeqa
Summing over $s_1,s_2,s_3$ we get
\beqa
\sum_r e^{2\pi i r/3} n_{r,t_1,t_2,t_3} = -4
\delta_{(t_1,t_2,t_3),(0,0,0)}\; .
\eeqa
Recalling the T-dual interpretation of $n_{r,t_1,t_2,t_3}$ as
multiplicities in $\g_{\theta,3',(t_1,t_2,t_3)}$, we obtain
\beqa
\Tr \g_{\theta,3',(t_1,t_2,t_3)} & = & -4 \delta_{(t_1,t_2,t_3),(0,0,0)}\; .
\eeqa
These are precisely the $\theta$-twisted tadpole conditions in the T-dual
model \cite{lykken}, i.e. the $\Omega R(-)^{F_L}$ orientifold of
$T^6/\IZ_3$. Namely the origin is fixed under the orientifold action and
receives a crosscap contribution equal to $-4$, while the remaining points
are not fixed under the orientifold action, and do not have such
contribution. Clearly, one can repeat the exercise for $\theta^2$, leading
to the analogous result.

\bigskip

To show the technique is completely general, let us repeat the exercise
for the $\IZ_6$ orientifold and its T-dual along the first two complex
planes, which is itself a $\IZ_6$ orientifold with D9- and D5$_3$-branes
mapping to D5$_3'$- and D9-branes.

We consider a quite general configuration with D9-branes with order
three Wilson lines along all three complex planes, and D5$_3$ branes
sitting at different $\theta^2$-fixed points in the first and second
plane, and with order three Wilson lines on the third. The tadpole
conditions are given in (\ref{tad1}). Let us concentrate on 
those associated with
$\theta^2$, which can be condensed as
\beqa
(m_1,m_2,m_3): & \;
\Tr (\gamma_{\theta^2,9}\gamma_{W_1,9}^{m_1} \gamma_{W_2,9}^{m_2}
\gamma_{W_3,9}^{2m_3})\;  
+ 3\Tr (\gamma_{\theta^2,5,(m_1,m_2)} \gamma_{W_3,5_3,(m_1,m_2)}^{2m_3}) = 
12 \delta_{m_1,0}\delta_{m_2,0}+ 4\; . 
\label{tadz6one}
\eeqa

The order three Wilson lines along the first complex plane have the
structure determined in Subsection~\ref{wl}A. In particular they commute
with $\gamma_{\theta^2,9}$ and among themselves (and so does 
$\gamma_{W_3,9}$), so we may diagonalize them simultaneously. Let us
denote by $n_{r,s_1,s_2,s_3}$ the number of eigenvalues $e^{2\pi ir/3}$,
$e^{2\pi i s_a/3}$, in $\gamma_{\theta^2,9}$, $\gamma_{W_a,9}$,
respectively. The matrix $\gamma_{W_3,5_5,(m_1,m_2)}$ also commutes with
$\gamma_{\theta^2,5_3,(m_1,m_2)}$, and we may diagonalize these as well.
Let us denote by $m_{rm_1m_2s_3}$ the number of eigenvalues $e^{2\pi
ir/3}$, $e^{2\pi i s_3/3}$ in $\gamma_{\theta^2,5_3,(m_1,m_2)}$,
$\gamma_{W_3,5_3,(m_1,m_2)}$, respectively.

Clearly, after T-dualizing along the first two complex planes D9-branes
with $W_1$, $W_2$ Wilson line eigenvalues determined by $s_1$, $s_2$ are
mapped to D5$'_3$-branes sitting at the point $(s_1,s_2)$. Similarly,
D5$_3$-branes at $(m_1,m_2)$ map to D9$'$-branes with $W_1$, $W_2$ Wilson
line eigenvalues determined by $m_1$, $m_2$. Hence, in the T-dual picture
$n_{rs_1s_2s_3}$ denotes the number of entries with eigenvalues 
$e^{2\pi i r/3}$, $e^{2\pi i s_3/3}$ in $\gamma_{\theta^2,5'_3,(s_1,s_2)}$, 
$\gamma_{W_3,5'_3,(s_1,s_2)}$, respectively, and $m_{rm_1m_2s_3}$ denotes 
the number of entries with eigenvalues $e^{2\pi i r/3}$, $e^{2\pi i m_1/3}$, 
$e^{2\pi i m_2/3}$, $e^{2\pi i s_3/3}$ in $\gamma_{\theta^2,9'}$,
$\gamma_{W_a,9'}$, respectively. This completely specifies the
T-dual model. Our purpose, in what follows, is to show that the consistency
conditions in the original model, after the discrete Fourier transform,
provide the consistency conditions in the T-dual picture.

The tadpole condition (\ref{tadz6one}) reads
\beqa
\sum_{r,s_1,s_2,s_3} e^{2\pi i \frac r3} e^{2\pi i \frac{ m_1s_1}{3}} 
e^{2\pi i \frac{m_2s_2}{3}} e^{2\pi i \frac{2m_3s_3}{3}} n_{rs_1s_2s_3} +
& 3 \sum_{r,s_3} e^{2\pi i\frac r3} e^{2\pi i \, \frac{2m_3s_3}{3}} 
m_{rm_1m_2m_3} =  & \nonumber \\
& = 12 \delta_{m_1,0}\delta_{m_2,0} + 4\; . &
\label{tadz6two}
\eeqa
Multiplying by $e^{-2\pi i\, m_1t_1/3} e^{-2\pi i\, m_2t_2/3}$ and summing 
over $m_1$, $m_2$, we get
\beqa
& 9 \sum_{r,s_1,s_2,s_3}\; e^{2\pi i\frac r3}\; \delta_{s_1,t_1} \;
\delta_{s_2,t_2} \; e^{2\pi i \frac{2m_3 s_3}{3}} n_{rs_1s_2s_3} + &
\nonumber \\
& + 3\sum_{rm_3} e^{2\pi i\frac{r}{3}}\; e^{-2\pi i\frac{m_1t_1}{3}}\; 
e^{-2\pi i\frac{m_2t_2}{3}}\; e^{2\pi i \frac{2m_3s_3}{3}}\;
m_{rm_1m_2m_3} & = 12 + 36 \, \delta_{t_1,0}\delta_{t_2,0} \; .
\nonumber
\eeqa
Summing over $s_1,s_2$ we get
\beqa
& \sum_{r,m_1,m_2,m_3} e^{2\pi i\frac r3}\; e^{-2\pi i\frac{m_1t_1}{3}}\; 
e^{-2\pi i\frac{m_2t_2}{3}}\; e^{2\pi i\frac{2m_3s_3}{3}}\; m_{rm_1m_2m_3}
+ & \nonumber \\
& + 3 \sum_{r,m_3} e^{2\pi ir/3}\; e^{2\pi i\, 2m_3s_3/3}\; n_{rt_1t_2m_3}
& = 12\, \delta_{t_1,0}\, \delta_{t_2,0}+4\; .
\label{tadz6three}
\eeqa
Now, using the T-dual interpretation of $n_{rt_1t_2s_3}$ and $m_{r m_1
m_2 m_3}$, we finally  obtain
\beqa
\Tr ( \g_{\theta^2,9'} \g_{W_1,9'}^{t_1} \g_{W_2,9'}^{t_2}
\g_{W_3,9'}^{2s_3}) + 3 \Tr (\g_{\theta^2,5'_3,(t_1,t_2)}
\g_{W_3,5'_3,(t_1,t_2)}^{2s_3}) \ =\ 12 \delta_{(t_1,t_2),(0,0)}+4\; ,
\eeqa
which is precisely the $\theta^2$ tadpole condition at the point
$(t_1,t_2,s_3)$, which is of the form (\ref{tadz6one}) since the T-dual picture is again an $\Omega$ orientifold of $T^6/\IZ_6$.

Notice that the T-duality discussed for the $\IZ_6$ orientifold in Section
\ref{tadpole} is along the third complex plane, therefore it is identical
to that performed for the $\IZ_3$ orientifold in this Appendix (and
differs from that studied here for the $\IZ_6$ case).

\end{document}